\documentstyle[floats,prl,aps]{revtex}
\begin{document}
\renewcommand{\thefootnote}{\fnsymbol{footnote}}
\draft
\title{\large\bf 
Integrable Kondo impurities in one-dimensional 
extended Hubbard models} 

\author { Huan-Qiang Zhou, Xiang-Yu Ge, Jon Links and Mark D. Gould } 

\address{      Department of Mathematics, University of Queensland,
		     Brisbane, Qld 4072, Australia}

\maketitle

\vspace{10pt}

\begin{abstract}
Three kinds of integrable Kondo problems in  one-dimensional extended
Hubbard models are studied by means of the boundary graded quantum inverse
scattering method.
The boundary $K$ matrices depending on the local moments of the impurities 
are presented  as a nontrivial realization of the graded reflection equation
algebras acting in a $(2 s_\alpha+1)$-dimensional impurity Hilbert space.
Further, these models are solved  using the algebraic Bethe ansatz method
and the Bethe ansatz equations are obtained. 
\end{abstract}




\def\a{\alpha}
\def\b{\beta}
\def\d{\delta}
\def\e{\epsilon}
\def\g{\gamma}
\def\k{\kappa}
\def\l{\lambda}
\def\o{\omega}
\def\t{\theta}
\def\s{\sigma}
\def\z{\zeta}
\def\D{\Delta}
\def\L{\Lambda}


\def\beq{\begin{equation}}
\def\eeq{\end{equation}}
\def\bea{\begin{eqnarray}}
\def\eea{\end{eqnarray}}
\def\ba{\begin{array}}
\def\ea{\end{array}}
\def\no{\nonumber}
\def\le{\langle}
\def\re{\rangle}
\def\lt{\left}
\def\rt{\right}

\newcommand{\reff}[1]{eq.~(\ref{#1})}

\vskip.3in

\section{Introduction \label{int}}

The study of integrable models of correlated electrons 
with open boundary conditions has been
the subject of considerable attention 
\cite{fk,Gon94a,acf,E,Zhou97,a,bf,BRA98,zz}. 
Recently it has become
apparent that for models on open chains it is possible to obtain
integrable impurity boundary conditions as operators which need
not be expressed in terms of the (super)symmetry of the bulk model. 
A very important application of this procedure is in the context of
Kondo; i.e. spin impurities in models of correlated electrons. For the
case of the supersymmetric $t-J$ model boundary spin-$\frac{1}{2}$ impurities were
introduced in \cite{hpw} and the resulting model solved by means of the
co-ordinate Bethe ansatz method. 

A reformulation of this model in the context of the Quantum Inverse
Scattering Method (QISM) was given in \cite{ZG} demonstrating that the
model could be obtained via a family of commuting transfer matrices and
thus establishing integrability. Central to this approach is the  
representations of the reflection equation algebras originally introduced
by Sklyanin \cite{Skl88}. Such a solution guarantees that boundary terms may
be applied to any model whose bulk integrability is associated with a
solution of the Yang-Baxter equation. An interesting observation made in
\cite{ZG} was that the necessary solution of the reflection equation was
not regular in the sense that it is not obtained by ``dressing''; i.e. 
it can not be factorized into a
product of local monodromy matrices and a $c$-number matrix.  

By utilizing the underlying algebraic structure it was subsequently
shown in \cite{zglg} that a more general classes of integrable $t-J$ models
with Kondo impurities exist. These were derived from both $gl(2|1)$ and
$gl(3)$ invariant solutions of the Yang-Baxter equation and the solution
of the reflection equation was extended to
accomodate arbitrary spin $s$ impurities situated on the boundaries.
Again, the new solutions of the reflection equation are not regular. 
Moreover it was also demonstrated in \cite{zglg} that the algebraic Bethe
ansatz is applicable for these models and explicit solutions were given.

Recently, the work of Frahm and Slavnov \cite{fs} has provided a
representation theoretic explanation for the existence of these
non-regular solutions of the relection equation. In essence, such
solutions are obtained by suitable projection onto a subspace of the
impurity Hilbert space for a regular solution. A consequence of this
projection method is that the remaining (super)symmetry in the  new 
boundary operator on the impurity site corresponds to a subalgebra oof
the (super)symmetry of the original regular solution. As examples, this
was illustrated in \cite{fs} for the case of $gl(m)$ impurites coupled to an
open $gl(n)$ invaraint chain for $m<n$ and a reproduction of the
integrable $t-J$ model with Kondo impurities given in \cite{zglg}. 

It is immediately evident in view of these results that integrable 
spin impurities,
being characterized by the simplest Lie algebra $su(2)$, can be readily
obtained from regular solutions coming from the larger (super)symmetry
associated with the model in the bulk. In particular, it is possible to
obtain integrable boundary Kondo impurity models associated with the Lie
algebra $gl(4)$ and superalgebras $gl(3|1)$ and $gl(2|2)$ which we
investigate here. In each case, the bulk Hamiltonian can be expressed in
the form of an extended Hubbard model and thus is worthy of
investigation in terms of the physical properties that are exhibited.
The bulk Hamiltonian associated with the $gl(2|2)$ solution is well
known from previous works of Essler et. al. \cite{eks}. However the other
two cases give rise to bulk Hamiltonians which are apparently new. 

In the next section we introduce the three forms of extended Hubbard
models with integrable boundary Kondo impurities. Following this we
undertake an algebraic Bethe ansatz approach to solve each case. In the
last section we conclude with some final remarks.

\section{Integrable non-$c$-number boundary $K$-matrices and Kondo impurities
in one-dimensional extended Hubbard models \label{Boun}}

Let $c_{j,\s}$ and $c_{j,\s}^{\dagger}$ denote fermionic creation and
annihilation operators for spin $\s$ at
site $j$, which satisfy the anti-commutation relations 
$\{c_{i,\s}^\dagger, c_{j,\tau}\}=\d_{ij}\d_{\s\tau}$, where 
$i,j=1,2,\cdots,L$ and $\s,\tau=\uparrow,\;\downarrow$. We consider 
the following Hamiltonian which describes two impurities coupled to 
the supersymmetric extended Hubbard open chain of Essler et. al.
\cite{eks},
\bea
H&=&-\sum _{j=1,\s}^{L-1}
(c_{j,\s}^\dagger c_{j+1,\s}+{\rm H.c.})
  (1-n_{j,-\s}-n_{j+1,-\s})
  \no\\
   & &-\sum ^{L-1}_{j=1}(c_{j,\uparrow}^\dagger c_{j,\downarrow}^\dagger
   c_{j+1,\downarrow}c_{j+1,\uparrow}
   +{\rm H.c}) 
   +2\sum ^{L-1}_{j=1}({\bf S}_j\cdot {\bf S}_{j+1}
 -\frac{1}{4}n_jn_{j+1})\no\\
 & & +J_a {\bf S}_1 \cdot {\bf S}_a +V_a n_1 +U_a n_{1\uparrow}
 n_{1\downarrow}
  +J_b {\bf S}_L \cdot {\bf S}_b +V_b n_L +U_b n_{L\uparrow}
 n_{L\downarrow},
  \label{ham1}
\eea
where $J_\a,V_\a $ and $U_\a (\a=a,b)$ 
are the Kondo coupling constants ,the
impurity scalar potentials and the boundary Hubbard-like interaction
constants,respectively; ${\bf S}$
is the vector spin operator
for the conduction electrons; ${\bf S}_{\a} (\a = a,b)$ are the local
moments with spin-$\frac {1}{2}$ located at the left and right ends of
the system respectively;
 $n_{j\s}$ is the number density operator
$n_{j\s}=c_{j\s}^{\dagger}c_{j\s}$,
$n_j=n_{j\uparrow}+n_{j\downarrow}$.

The supersymmetry algebra underlying the bulk
Hamiltonian of this model is
$gl(2|2)$. 
 It is quite interesting to note that although
the introduction  of the impurities spoils the supersymmetry, there 
still remains  $u (2)\otimes u(2)$ symmetry in the Hamiltonian (\ref {ham1})
whose representation contains the spin and $\eta$-pairing realizations.
As a result, one may add some terms like $U \sum _{j=1}^L n_{j\uparrow}n_{j\downarrow}$,$\mu \sum ^L_{j=1}n_j$ and $h \sum _{j=1}^L
(n_{j\uparrow}-n_{j\downarrow})$ to the Hamiltonian (\ref {ham1}),without
spoiling the integrability. Below we will establish the quantum integrability of the Hamiltonian
(\ref{ham1}) for a special choice of the model parameters $J_\a$,
$V_\a$,and $U_\a$
\beq
J_\a = -\frac {2}{c_\a(c_\a+2s_\a+1)},
V_\a = -\frac {c_\a^2+2c_\a s_\a-s_\a}{c_\a(c_\a+2s_\a+1)},
U_\a = -\frac {2s_\a-c^2_\a-c_\a(2s_\a-1)}{c_\a(c_\a+2s_\a+1)}. 
\eeq
This is achieved by showing that it can be derived from
the (graded) boundary quantum inverse scattering method \cite
{Zhou97,BRA98}.Here we emphasize that a special case of this
model,corresponding to  $s_\a = \frac{1}{2}$, has been studied in  
\cite {ZGG98}.

The second choice of couplings which leads to an integrable model 
is given by 
\bea
H&=&-\sum _{j=1,\s}^{L-1}
(c_{j,\s}^\dagger c_{j+1,\s}+{\rm H.c.})
  (1-n_{j,-\s}-n_{j+1,-\s})
  \no\\
   & &-\sum ^{L-1}_{j=1}(c_{j,\uparrow}^\dagger c_{j,\downarrow}^\dagger
   c_{j+1,\downarrow}c_{j+1,\uparrow}
   +{\rm H.c}) 
   -2\sum ^{L-1}_{j=1}({\bf S}_j\cdot {\bf S}_{j+1}
   +\frac{3}{4}n_jn_{j+1})\no\\
   & &-2\sum^{L-1}_{j=1}n_{j,\downarrow}n_{j,\uparrow} 
   (n_{j+1,\downarrow}n_{j+1,\uparrow}-n_{j+1}) 
   -2\sum^{L-1}_{j=1}n_{j+1,\downarrow}n_{j+1,\uparrow} 
   (n_{j,\downarrow}n_{j,\uparrow}-n_{j})\no\\ 
 & & +J_a {\bf S}_1 \cdot {\bf S}_a +V_a n_1 +U_a n_{1\uparrow}
 n_{1\downarrow}
  +J_b {\bf S}_L \cdot {\bf S}_b +V_b n_L +U_b n_{L\uparrow}
 n_{L\downarrow},
  \label{ham2}
\eea
In this case we can introduce integrable Kondo impurities on the
boundary by choosing
\beq
J_\a = \frac {8}{(2c_\a+2s_\a+1)(2c_\a-2s_\a-1)},
V_\a = -\frac {4c_\a^2+4c_\a-4s_\a(s_\a+1)-3}
      {(2c_\a+2s_\a+1)(2c_\a-2s_\a-1)},
U_\a = \frac {4c^2_\a+8c_\a-4s_\a(s_\a+1)-5}
      {(2c_\a+2s_\a+1)(2c_\a-2s_\a-1)}. 
\label{con2} \eeq

A third choice of couplings which leads to an integrable model is 
\bea
H&=&-\sum _{j=1,\s}^{L-1}
(c_{j,\s}^\dagger c_{j+1,\s}+{\rm H.c.})
  (1-n_{j,-\s}-n_{j+1,-\s})
  \no\\
   & &-\sum ^{L-1}_{j=1}(c_{j,\uparrow}^\dagger c_{j,\downarrow}^\dagger
   c_{j+1,\downarrow}c_{j+1,\uparrow}
   +{\rm H.c}) 
   -2\sum ^{L-1}_{j=1}({\bf S}_j\cdot {\bf S}_{j+1}
 -\frac{1}{4}n_jn_{j+1})
   -2\sum^{L-1}_{j=1}n_{j,\downarrow}n_{j,\uparrow} 
   n_{j+1,\downarrow}n_{j+1,\uparrow}\no\\ 
 & & +J_a {\bf S}_1 \cdot {\bf S}_a +V_a n_1 +U_a n_{1\uparrow}
 n_{1\downarrow}
  +J_b {\bf S}_L \cdot {\bf S}_b +V_b n_L +U_b n_{L\uparrow}
 n_{L\downarrow},
  \label{ham3}
\eea
where integrable Kondo impurities on the
boundary are obtained by the choice 
\beq
J_\a = \frac {8}{(2c_\a+2s_\a+1)(2c_\a-2s_\a-1)},
V_\a = \frac {(2c_\a^2-1)^2-4s_\a(s_\a+1)}
      {(2c_\a+2s_\a+1)(2c_\a-2s_\a-1)},
U_\a = -\frac {4(c^2_\a-1)^2-(2s_\a+1)^2}
      {(2c_\a+2s_\a+1)(2c_\a-2s_\a-1)}. 
\label{con3} \eeq

Let us recall that the Hamiltonian of
the 1D  supersymmetric extended Hubbard  model with periodic
boundary conditions
commutes with the transfer matrix, which is the supertrace of the
monodromy matrix $T(u)$
\beq
T(u) = R_{0L}(u)\cdots R_{01}(u). \label{matrix-t}
\eeq
Here the quantum R-matrix 
$ R(u)$ comes from the fundamental representation of $gl(2|2)$  and  
takes the form 
\beq
R(u)=\left( \begin {array} {cccccccccccccccc}
u-2&0&0&0&0&0&0&0&0&0&0&0&0&0&0&0\\
0&u&0&-2&0&0&0&0&0&0&0&0&0&0&0&0\\
0&0&u&0&0&0&0&0&-2&0&0&0&0&0&0&0\\
0&0&0&u&0&0&0&0&0&0&0&0&-2&0&0&0\\
0&-2&0&0&u&0&0&0&0&0&0&0&0&0&0&0\\
0&0&0&0&0&u-2&0&0&0&0&0&0&0&0&0&0\\
0&0&0&0&0&0&u&0&0&-2&0&0&0&0&0&0\\
0&0&0&0&0&0&0&u&0&0&0&0&0&-2&0&0\\
0&0&-2&0&0&0&0&0&u&0&0&0&0&0&0&0\\
0&0&0&0&0&0&-2&0&0&u&0&0&0&0&0&0\\
0&0&0&0&0&0&0&0&0&0&u+2&0&0&0&0&0\\
0&0&0&0&0&0&0&0&0&0&0&u&0&0&2&0\\
0&0&0&-2&0&0&0&0&0&0&0&0&u&0&0&0\\
0&0&0&0&0&0&0&-2&0&0&0&0&0&u&0&0\\
0&0&0&0&0&0&0&0&0&0&0&2&0&0&u&0\\
0&0&0&0&0&0&0&0&0&0&0&0&0&0&0&u+2
\end {array}  \right ), \label {r1}
\eeq
It should be noted that the supertrace
is carried out for the auxiliary superspace $V$.
The elements of the supermatrix $T(u)$ are the generators
of an associative superalgebra ${\cal A}$ defined by the relations
\beq
R_{12}(u_1-u_2) \stackrel {1}{T}(u_1) \stackrel {2}{T}(u_2) =
   \stackrel {2}{T}(u_2) \stackrel {1}{T}(u_1)R_{12}(u_1-u_2),\label{rtt-ttr} 
\eeq
where $\stackrel {1}{X} \equiv  X \otimes 1,~
\stackrel {2}{X} \equiv  1 \otimes X$
for any supermatrix $ X \in End(V) $. For later use, we list some useful
properties enjoyed by the R-matrix:
(i) Unitarity:   $  R_{12}(u)R_{21}(-u) = \rho (u)$ and (ii)
 Crossing-unitarity:  $  R^{st_2}_{12}(-u+1)R^{st_2}_{21}(u) =
         \tilde {\rho }(u)$
with $\rho (u),\tilde \rho (u)$ being some  scalar functions.

In order to describe integrable  models on  open
chains, we introduce two associative
superalgebras ${\cal T}_-$  and ${\cal T}_+$ defined by the R-matrix
$R(u_1-u_2)$ and the relations \cite {Zhou97,BRA98} 
\beq
R_{12}(u_1-u_2)\stackrel {1}{\cal T}_-(u_1) R_{21}(u_1+u_2)
  \stackrel {2}{\cal T}_-(u_2)
=  \stackrel {2}{\cal T}_-(u_2) R_{12}(u_1+u_2)
  \stackrel {1}{\cal T}_-(u_1) R_{21}(u_1-u_2),  \label{reflection1}
\eeq
\bea
&&R_{21}^{st_1 ist_2}(-u_1+u_2)\stackrel {1}{{\cal T}^{st_1}_+}
  (u_1) \{\;[\;R^{st_1}_{21}(u_1+u_2)\;]^{-1}\;\}^{ist_2}
  \stackrel {2}{{\cal T}^{ist_2}_+}(u_2)\no\\
&&~~~~~~~~~~~~~~~=\stackrel {2}{{\cal T}^{ist_2}_+}(u_2)
  \{\;[\;R^{ist_2}_{12}(u_1+u_2)\;]^{-1}\;\}^{st_1}
  \stackrel {1}{{\cal T}^{st_1}_+}(u_1) R_{12}^{st_1 ist_2}(-u_1+u_2),
  \label{reflection2}
\eea
respectively. Here the supertransposition $st_{\alpha}~(\alpha =1,2)$ 
is only carried out in the
$\alpha$-th factor superspace of $V \otimes V$, whereas $ist_{\alpha}$ denotes
the inverse operation of  $st_{\alpha}$. By modifying Sklyanin's 
arguments \cite{Skl88}, one
may show that the quantities $\tau(u)$ given by
$\tau (u) = str ({\cal T}_+(u){\cal T}_-(u))$
constitute a commutative family, i.e.,
        $[\tau (u_1),\tau (u_2)] = 0$. 

One can obtain a class of realizations of the superalgebras ${\cal T}_+$  and
${\cal T}_-$  by choosing  ${\cal T}_{\pm}(u)$ to be the form
\beq
{\cal T}_-(u) = T_-(u) \tilde {\cal T}_-(u) T^{-1}_-(-u),~~~~~~ 
{\cal T}^{st}_+(u) = T^{st}_+(u) \tilde {\cal T}^{st}_+(u) 
  \lt(T^{-1}_+(-u)\rt)^{st}\label{t-,t+} 
\eeq
with
\beq
T_-(u) = R_{0M}(u) \cdots R_{01}(u),~~~~
T_+(u) = R_{0L}(u) \cdots R_{0,M+1}(u),~~~~ 
\tilde {\cal T}_{\pm}(u) = K_{\pm}(u),
\eeq
where $K_{\pm}(u)$, called boundary K-matrices, 
are representations of  ${\cal T}_{\pm}  $ in some representation
superspace. 

We now solve (\ref{reflection1}) and (\ref{reflection2}) 
for $K_-(u)$ and $K_+(u)$. For the quantum $R$-matrix (\ref {r1}),
one may 
check that the matrix $K_-(u)$ given by
\beq
K_-(u)= \left ( \begin {array}{cccc}
1&0&0&0\\
0&1&0&0\\
0&0&A_-(u)&B_-(u)\\
0&0&C_-(u)&D_-(u)
\end {array} \right ),\label{k-1}
\eeq
where
\bea
A_-(u)&=&-\frac{u^2+2u-4c^2_a-4c_a(2s_a+1)+4u{\bf S}^z_a}
{(u-2c_a)(u-2c_a-4s_a-2)},\no\\
B_-(u)&=&-\frac{4u{\bf S}^-_a}
{(u-2c_a)(u-2c_a-4s_a-2)},\no\\
C_-(u)&=&-\frac{4u{\bf S}^+_a}
{(u-2c_a)(u-2c_a-4s_a-2)},\no\\
D_-(u)&=&-\frac{u^2+2u-4c^2_a-4c_a(2s_a+1)-4u{\bf S}^z_a}
{(u-2c_a)(u-2c_a-4s_a-2)},
\eea
satisfies (\ref{reflection1}). Here ${\bf S}^{\pm}={\bf S}^x \pm
i{\bf S}^y$.
The matrix $K_+(u)$ can be obtained from the isomorphism of the
superalgebras  ${\cal T}_-  $ and ${\cal T}_+  $. Indeed, given a solution
${\cal T}_- $ of (\ref{reflection1}), then ${\cal T}_+(u)$ defined by
\beq
{\cal T}_+^{st}(u) =  {\cal T}_-(-u)\label{t+t-1}
\eeq
is a solution of (\ref{reflection2}). 
The proof follows from some algebraic computations upon
substituting (\ref{t+t-1}) into  
(\ref{reflection2}) and making use
of the properties of the R-matrix .
Therefore, one may choose the boundary matrix $K_+(u)$ as 
\beq
K_+(u)=\left ( \begin {array} {cccc}
1&0&0&0\\
0&1&0&0\\
0&0& A_+(u)&B_+(u)\\
0&0&C_+(u)& D_+(u)
\end {array} \right )\label{k^I+1}
\eeq
with
\bea
A_+(u)&=&-\frac{u^2-2u-4c^2_b-4c_b(2s_b-1)+8s_b+4u{\bf S}^z_b}
{(u-2c_b+2)(u-2c_b-4s_b)},\no\\
B_+(u)&=&-\frac{4u{\bf S}^-_b}{(u-2c_b+2)(u-2c_b-4s_b)},\no\\
C_+(u)&=&-\frac{4u{\bf S}^+_b}{(u-2c_b+2)(u-2c_b-4s_b)},\no\\
D_+(u)&=&-\frac{u^2-2u-4c^2_b-4c_b(2s_b-1)+8s_b-4u{\bf S}^z_b}
{(u-2c_b+2)(u-2c_b-4s_b)},
\eea

Now it can be shown  that 
Hamiltonian (\ref{ham1}) is related to the second derivative of the
boundary transfer matrix
$\tau (u)$ with respect to the spectral parameter $u$ at $u=0$ (up to an unimportant additive constant)
\bea
H&=&\frac {\tau'' (0)}{4(V+2W)}=
  \sum _{j=1}^{L-1} h_{j,j+1} + \frac {1}{2} \stackrel {1}{K'}_-(0)
+\frac {1}{2(V+2W)}\lt[str_0\lt(\stackrel {0}{K}_+(0)G_{L0}\rt)\rt.\no\\
& &\lt.+2\,str_0\lt(\stackrel {0}{K'}_+(0)H_{L0}^R\rt)+
  str_0\lt(\stackrel {0}{K}_+(0)\lt(H^R_{L0}\rt)^2\rt)\rt],\label{derived-h}
\eea
with
$$h=-\frac{1}{2}\frac{d}{du}PR(u) $$
where $P$ denotes the graded
permutation operator,
and the subscript $0$ denotes the 4-dimensional auxiliary superspace $V=C^{2,2}$ with
the grading $P[i]=0~ {\rm if} ~i=1,2$ and $1~ {\rm if}~ i=3,4$, and
\bea
V&=&str_0 K'_+(0),~~~~~~W=str_0 \lt(\stackrel {0}{K}_+(0)
H_{L0}^R\rt),\no\\   
H^R_{i,j}&=&P_{i,j}R'_{i,j}(0),~~~~~G_{i,j}=P_{i,j}R''_{i,j}(0).
\eea
This implies that this model, as with the following two model we will
study, admits an infinite number
of mutually commuting conserved currents, thus
assuring its integrability.
 
The second choice of integrable couplings results from use of an
$R$-matrix obtained by imposing ${\bf Z}_2$ grading  associated with two
bosonic and two fermionic states to the fundamental 
$su(4)$ $R$-matrix which reads
\beq
R(u)=\left( \begin {array} {cccccccccccccccc}
u-2&0&0&0&0&0&0&0&0&0&0&0&0&0&0&0\\
0&u&0&-2&0&0&0&0&0&0&0&0&0&0&0&0\\
0&0&u&0&0&0&0&0&-2&0&0&0&0&0&0&0\\
0&0&0&u&0&0&0&0&0&0&0&0&-2&0&0&0\\
0&-2&0&0&u&0&0&0&0&0&0&0&0&0&0&0\\
0&0&0&0&0&u-2&0&0&0&0&0&0&0&0&0&0\\
0&0&0&0&0&0&u&0&0&-2&0&0&0&0&0&0\\
0&0&0&0&0&0&0&u&0&0&0&0&0&-2&0&0\\
0&0&-2&0&0&0&0&0&u&0&0&0&0&0&0&0\\
0&0&0&0&0&0&-2&0&0&u&0&0&0&0&0&0\\
0&0&0&0&0&0&0&0&0&0&-u+2&0&0&0&0&0\\
0&0&0&0&0&0&0&0&0&0&0&-u&0&0&2&0\\
0&0&0&-2&0&0&0&0&0&0&0&0&u&0&0&0\\
0&0&0&0&0&0&0&-2&0&0&0&0&0&u&0&0\\
0&0&0&0&0&0&0&0&0&0&0&2&0&0&-u&0\\
0&0&0&0&0&0&0&0&0&0&0&0&0&0&0&-u+2
\end {array}  \right ), \label {r2}
\eeq

We now solve (\ref{reflection1}) and (\ref{reflection2}) 
for $K_-(u)$ and $K_+(u)$. For  (\ref {r2}),
 we find  that the matrix $K_-(u)$ given by
\beq
K_-(u)= \left ( \begin {array}{cccc}
1&0&0&0\\
0&1&0&0\\
0&0&A_-(u)&B_-(u)\\
0&0&C_-(u)&D_-(u)
\end {array} \right ),\label{k-2}
\eeq
where
\bea
A_-(u)&=&-\frac{u^2-2u-4c^2_a+4s_a(s_a+1)+1-4u{\bf S}^z_a}
{(u+2c_a-2s_a-1)(u+2c_a+2s_a+1)},\no\\
B_-(u)&=&\frac{4u{\bf S}^-_a}
{(u+2c_a-2s_a-1)(u+2c_a+2s_a+1)},\no\\
C_-(u)&=&\frac{4u{\bf S}^+_a}
{(u+2c_a-2s_a-1)(u+2c_a+2s_a+1)},\no\\
D_-(u)&=&-\frac{u^2-2u-4c^2_a+4s_a(s_a+1)+1+4u{\bf S}^z_a}
{(u+2c_a-2s_a-1)(u+2c_a+2s_a+1)},
\eea
satisfies (\ref{reflection1}). 
The matrix $K_+(u)$ can again be obtained from the isomorphism of the
superalgebras  ${\cal T}_-  $ and ${\cal T}_+  $ through 
\beq
{\cal T}_+^{st}(u) =  {\cal T}_-(-u+4)\label{t+t-2}
.\eeq
Therefore, one choose the boundary matrix $K_+(u)$ as 
\beq
K_+(u)=\left ( \begin {array} {cccc}
1&0&0&0\\
0&1&0&0\\
0&0& A_+(u)&B_+(u)\\
0&0&C_+(u)& D_+(u)
\end {array} \right )\label{k^I+2}
\eeq
with
\bea
A_+(u)&=&\frac{u^2-6u-4c^2_b-8c_b+4s_a(s_b+1)+5-4(u-4){\bf S}^z_b}
{(u+2c_b-2s_b-3)(u+2c_b+2s_b-1)},\no\\
B_+(u)&=&-\frac{4(u-4){\bf S}^-_b}
{(u+2c_b-2s_b-3)(u+2c_b+2s_b-1)},\no\\
C_+(u)&=&-\frac{4(u-4){\bf S}^+_b}
{(u+2c_b-2s_b-3)(u+2c_b+2s_b-1)},\no\\
D_+(u)&=&\frac{u^2-6u-4c^2_b-8c_b+4s_a(s_b+1)+5+4(u-4){\bf S}^z_b}
{(u+2c_b-2s_b-3)(u+2c_b+2s_b-1)}.
\eea

For this example it can be shown  that the
Hamiltonian (\ref{ham2}) is related to the logarithmic derivative of the
transfer matrix
$\tau (u)$ with respect to the spectral parameter $u$ at $u=0$ (up to
an additive chemical potential term)
\beq
H= \sum _{j=1}^{L-1} h_{j,j+1} + \frac {1}{2} \stackrel {1}{K'}_-(0)
+\frac {str_0 K_+(0)H_{L0}}{str_0 K_+(0)},
\label{HAM} \eeq
with
$$h=-\frac{1}{2}\frac{d}{du}PR(u) $$
and subject to the constraints
(\ref{con2}). 

The third choice of integrable couplings results from use of the  
$R$-matrix obtained by imposing ${\bf Z}_2$ grading to the fundamental 
$gl(3|1)$ $R$-matrix  which reads
\beq
R(u)=\left( \begin {array} {cccccccccccccccc}
-u-2&0&0&0&0&0&0&0&0&0&0&0&0&0&0&0\\
0&u&0&-2&0&0&0&0&0&0&0&0&0&0&0&0\\
0&0&u&0&0&0&0&0&-2&0&0&0&0&0&0&0\\
0&0&0&u&0&0&0&0&0&0&0&0&-2&0&0&0\\
0&-2&0&0&u&0&0&0&0&0&0&0&0&0&0&0\\
0&0&0&0&0&u-2&0&0&0&0&0&0&0&0&0&0\\
0&0&0&0&0&0&u&0&0&-2&0&0&0&0&0&0\\
0&0&0&0&0&0&0&u&0&0&0&0&0&-2&0&0\\
0&0&-2&0&0&0&0&0&u&0&0&0&0&0&0&0\\
0&0&0&0&0&0&-2&0&0&u&0&0&0&0&0&0\\
0&0&0&0&0&0&0&0&0&0&-u+2&0&0&0&0&0\\
0&0&0&0&0&0&0&0&0&0&0&-u&0&0&2&0\\
0&0&0&-2&0&0&0&0&0&0&0&0&u&0&0&0\\
0&0&0&0&0&0&0&-2&0&0&0&0&0&u&0&0\\
0&0&0&0&0&0&0&0&0&0&0&2&0&0&-u&0\\
0&0&0&0&0&0&0&0&0&0&0&0&0&0&0&-u+2
\end {array}  \right ), \label {r3}
\eeq

Again we solve (\ref{reflection1}) and (\ref{reflection2}) 
for $K_-(u)$ and $K_+(u)$. For (\ref {r3})
we obtain 
\beq
K_-(u)= \left ( \begin {array}{cccc}
1&0&0&0\\
0&1&0&0\\
0&0&A_-(u)&B_-(u)\\
0&0&C_-(u)&D_-(u)
\end {array} \right ),\label{k-3}
\eeq
where
\bea
A_-(u)&=&-\frac{u^2-2u-4c^2_a+4s_a(s_a+1)+1-4u{\bf S}^z_a}
{(u+2c_a-2s_a-1)(u+2c_a+2s_a+1)},\no\\
B_-(u)&=&\frac{4u{\bf S}^-_a}
{(u+2c_a-2s_a-1)(u+2c_a+2s_a+1)},\no\\
C_-(u)&=&\frac{4u{\bf S}^+_a}
{(u+2c_a-2s_a-1)(u+2c_a+2s_a+1)},\no\\
D_-(u)&=&-\frac{u^2-2u-4c^2_a+4s_a(s_a+1)+1+4u{\bf S}^z_a}
{(u+2c_a-2s_a-1)(u+2c_a+2s_a+1)}.
\eea
and 
\beq
{\cal T}_+^{st}(u) = J {\cal T}_-(-u+2)
,~~~~~~~~J=diag(1,-1,1,1),
\label{t+t-3}
\eeq
giving 
\beq
K_+(u)=\left ( \begin {array} {cccc}
1&0&0&0\\
0&-1&0&0\\
0&0& A_+(u)&B_+(u)\\
0&0&C_+(u)& D_+(u)
\end {array} \right )\label{k^I+3}
\eeq
with
\bea
A_+(u)&=&-\frac{u^2-2u-4c^2_b+4s_a(s_b+1)+1-4(u-2){\bf S}^z_b}
{(u+2c_b-2s_b-3)(u+2c_b+2s_b-1)},\no\\
B_+(u)&=&\frac{4(u-2){\bf S}^-_b}
{(u+2c_b-2s_b-3)(u+2c_b+2s_b-1)},\no\\
C_+(u)&=&\frac{4(u-2){\bf S}^+_b}
{(u+2c_b-2s_b-3)(u+2c_b+2s_b-1)},\no\\
D_+(u)&=&-\frac{u^2-2u-4c^2_b+4s_a(s_b+1)+1+4(u-2){\bf S}^z_b}
{(u+2c_b-2s_b-3)(u+2c_b+2s_b-1)}.
\eea

The Hamiltonian (\ref{ham3}) is related to the logarithmic derivative of the
transfer matrix
$\tau (u)$ with respect to the spectral parameter $u$ at $u=0$ (up to
an additive chemical potential term)
\beq
H= \sum _{j=1}^{L-1} h_{j,j+1} + \frac {1}{2} \stackrel {1}{K'}_-(0)
+\frac {str_0 K_+(0)H_{L0}}{str_0 K_+(0)},
\label{HAM} \eeq
with
$$h=-\frac{1}{2}\frac{d}{du}PR(u). $$
For this case we obtain (\ref{ham3}) subject to the constraints
(\ref{con3}).

\section{The Bethe ansatz solutions \label{bethe}}

Having established the quantum integrability of the models, let us now
diagonalize the Hamiltonians 
by means of the algebraic Bethe ansatz method
\cite {Skl88,Gon94}.
For the first case (\ref{ham1}), introduce the `doubled' monodromy matrix $U(u)$
\beq
U(u)=T(u)K_-(u)\tilde{T}(u) \equiv
 \left ( \begin {array}
{cccc}
{\cal A}(u)&{\cal B}_{1}(u)&{\cal B}_2(u)&{\cal B}_3(u)\\
{\cal C}_{1}(u)&{\cal D}_{11}(u)&{\cal D}_{12}(u)&{\cal D}_{13}(u)\\
{\cal C}_{2}(u)&{\cal D}_{21}(u)&{\cal D}_{22}(u)&{\cal D}_{23}(u)\\
{\cal C}_{3}(u)&{\cal D}_{31}(u)&{\cal D}_{32}(u)&{\cal D}_{33}(u)\\
\end {array} \right ).
\eeq
where $\tilde {T}(u)=T^{-1}(-u)$. Substituting into the reflection equation
(\ref {reflection1}) we may draw the following
commutation relations,
\bea
{\check {\cal D}}_{bd}(u_1){\cal B}_c(u_2)&=&\frac {(u_1-u_2-2)(u_1+u_2-4)}
{(u_1-u_2)(u_1+u_2-2)}r(u_1+u_2-2)^{eb}_{gh}r(u_1-u_2)^{ih}_{cd}
{\cal B}_e(u_2){\check {\cal D}}_{gi}(u_1)-\no\\
& &\frac {2(u_1-2)u_2}{(u_1+u_2-2)(u_1-1)(u_2-1)}r(2u_1-2)^{gb}_{cd}
{\cal B}_g(u_1) {\cal A}(u_2) + \no\\
& & \frac {2(u_1-2)}{(u_1-u_2)(u_1-1)}
r(2u_1-2)^{gb}_{id} {\cal B}_g (u_1) {\check {\cal D}}_{ic}(u_2),\label
{cr1}\\
{\cal A}(u_1){\cal B}_{\b}(u_2) 
&=&\frac {(u_1-u_2+2)(u_1+u_2)} {(u_1-u_2)(u_1+u_2-2)}
{\cal B}_{\b}(u_2){\cal A}(u_1)-\frac {2(u_1+u_2)}{(u_1-u_2)(u_1+u_2-2)}
{\cal B}_{\b}(u_1)A(u_2)\no\\
& &  +\frac {2}{u_1+u_2-2}[{\cal B}_{\a}(u_1)({\check {\cal D}}_{\a \b}(u_2)
-\frac{1}{u_2-1}\d _{\a \b}{\cal A}(u_2)].
\eea
Here $ {\cal D} _{bd}(u) = {\check {\cal D}}_{bd}(u) - \frac {1}{u-1}
\delta _{bd}{\cal A}(u)$ and the matrix $r(u)$, 
which in turn satisfies the quantum Yang-Baxter
equation, takes the form,
\bea
r^{11}_{11}(u)&=&1,~~~~~r^{22}_{22}(u)=r^{33}_{33}(u)=-\frac {u+2}{u-2},\no\\
 r^{12}_{12}(u)&=&r^{13}_{13}(u)=
r^{21}_{21}(u)=r^{31}_{31}(u)=
r^{23}_{23}(u)=r^{32}_{32}(u)=-\frac {2}{u-2},\no\\
r^{12}_{21}(u)&=&r^{21}_{12}(u)=
r^{13}_{31}(u)=r^{31}_{13}(u)=
\frac {u}{u-2},\no\\
r^{23}_{32}(u)&=&r^{32}_{23}(u)=
-\frac {u}{u-2}.
\eea
Next choose  Bethe state $|\Omega \rangle $ of the form 
\beq
|\Omega \rangle = {\cal B}_{i_1}(u_1) \cdots {\cal
B}_{i_N}(u_N)|0\rangle F^{i_1\cdots i_N},
\eeq
with $|0\rangle $ being the pseudovacuum. Acting the transfer
matrix $\tau (u)$
on the state $|\Omega\rangle$ we have
$\tau (u) |\Omega \rangle =\Lambda(u) |\Omega \rangle$ with the
eigenvalue
\bea
\Lambda (u)&=& \frac {u}{u-1}\frac {(c_b-\frac {u}{2})}{(c_b-\frac
{u}{2}-1)}\cdot
\frac {(c_b-\frac {u}{2}+2s_b+1)}{(c_b-\frac
{u}{2}+2s_b)}
\prod ^N_{j=1} \frac {(u+u_j)(u-u_j+2)}{(u-u_j)(u+u_j-2)}\no\\
& &+\frac {u}{u-1} (\frac {u}{u-2})^{2L} 
\prod ^N_{j=1} \frac {(u-u_j-2)(u+u_j-4)}{(u-u_j)(u+u_j-2)}
\Lambda ^{(1)}(u;\{u_i\}),
\eea
provided the parameters $\{ u_j\}$ satisfy
\beq
\frac {u_j}{u_j-2}\frac {(c_b-\frac {u_j}{2})}{(c_b-\frac
{u_j}{2}-1)}\cdot
\frac {(c_b-\frac {u_j}{2}+2s_b+1)}{(c_b-\frac
{u_j}{2}+2s_b)}
(\frac {u_j-2}{u_j})^{2L}=
\prod ^N_{\stackrel {i=1}{i \neq j}} \frac {(u_j-u_i-2)}{(u_j-u_i+2)}
\frac {(u_j+u_i-4)}{(u_j+u_i)}
\Lambda
^{(1)}(u_j;\{u_i\}). \label {bethe1-1}
\eeq
Here $\Lambda ^{(1)}(u;\{u_i\})$ is the eigenvalue of the transfer
matrix $\tau ^{(1)}(u)$ for the reduced problem which arises out of the
$r$ matrices from the first term in the right hand side of (\ref {cr1})
with
the reduced boundary $K$ matrices $K_{\pm}^{(1)}(u)$ 
\beq
K^{(1)}_-(u)=
  \left ( \begin {array}
{ccc}
1&0&0\\
0&A^{(1)}_-(u)&B^{(1)}_-(u)\\
0&C^{(1)}_-(u)&D^{(1)}_-(u)
\end {array} \right ),\label{k1-1}
\eeq
where
\bea
A^{(1)}_-(u)&=&-\frac {u^2-4c_a^{2}-8s_ac_a+4s_a+4(u-1) {\bf S}^z_a}
{(u-2c_a)(u-2c_a-4s_a-2)},\no\\
B^{(1)}_-(u)&=&-\frac {4(u-1) {\bf S}^-_a}
{(u-2c_a)(u-2c_a-4s_a-2)},\no\\
C^{(1)}_-(u)&=&-\frac {4(u-1) {\bf S}^+_a}
{(u-2c_a)(u-2c_a-4s_a-2)},\no\\
D^{(1)}_-(u)&=&-\frac {u^2-4c_a^{2}-8s_ac_a+4s_a-4(u-1) {\bf S}^z_a}
{(u-2c_a)(u-2c_a-4s_a-2)}.\no\\
\eea
and
\beq
K^{(1)}_+(u)=   \left ( \begin {array}
{ccc}
1&0&0\\
0&A^{(1)}_+(u)&B^{(1)}_+(u)\\
0&C^{(1)}_+(u)&D^{(1)}_+(u)
\end {array} \right ),
\eeq
where
\bea
A^{(1)}_+(u)&=&-\frac{u^2-2u-4c^2_b-4c_b(2s_b-1)+8s_b+4u{\bf S}^z_b}
{(u-2c_b+2)(u-2c_b-4s_b)},\no\\
B^{(1)}_+(u)&=&-\frac{4u{\bf S}^-_b}{(u-2c_b+2)(u-2c_b-4s_b)},\no\\
C^{(1)}_+(u)&=&-\frac{4u{\bf S}^+_b}{(u-2c_b+2)(u-2c_b-4s_b)},\no\\
D^{(1)}_+(u)&=&-\frac{u^2-2u-4c^2_b-4c_b(2s_b-1)+8s_b-4u{\bf S}^z_b}
{(u-2c_b+2)(u-2c_b-4s_b)}.  
\eea
Here $K^{(1)}_- (u)$, the boundary $K$ matrix after  the first
nesting, follows from the relation
\bea
{\check {\cal D}}_{dd}(u)|\Psi\rangle&\equiv&
\frac {u}{u-1}K^{(1)}_{dd}(u)|\Psi \rangle =
(K_-(u)_{dd}+\frac{1}{u-1})
(\frac{u}{u-2})^{2L}|\Psi\rangle,\no\\
{\check {\cal D}}_{db}(u)|\Psi\rangle&\equiv&
\frac {u}{u-1}K^{(1)}_{db}(u)|\Psi \rangle=
K_-(u)_{db} (\frac{u}{u-2})^{2L}|\Psi\rangle.
\eea
Indeed, applying the monodromy matrix
$T(u)$ and its ``adjoint'' ${\tilde T}(u)$ to the pseudovacuum, we have
\bea
T_{11}(u)|\Psi\rangle&=&|\Psi\rangle,~~~
T_{dd}(u)|\Psi\rangle=(\frac{u}{u-2})^L|\Psi\rangle,\no\\
T_{1d}(u)|\Psi\rangle&\neq& 0,~~~
T_{db}(u)|\Psi\rangle=0,~~~
T_{d1}(u)|\Psi\rangle= 0,\no\\
\tilde{T}_{11}(u)|\Psi\rangle&=&|\Psi\rangle,~~~
\tilde{T}_{dd}(u)|\Psi\rangle=(\frac{u}{u-2})^L|\Psi\rangle,\no\\
\tilde{T}_{1d}(u)|\Psi\rangle&\neq& 0,~~~
\tilde{T}_{db}(u)|\Psi\rangle=0,~~~
\tilde{T}_{d1}(u)|\Psi\rangle=0.
\eea
where $d\neq b,~~~ d,b=2,3,4$. Then we have
\bea
{\cal A}(u)|\Psi\rangle&=&|\Psi\rangle,\no\\
{\cal B}_d(u)|\Psi\rangle&\neq&0,~~~~~~
{\cal C}_d(u)|\Psi\rangle= 0,\no\\
{\cal D}_{db}(u)|\Psi\rangle&=&
(\frac{u}{u-2})^{2L} K_-(u)_{db}|\Psi\rangle,\no\\
{\cal D}_{dd}(u)|\Psi\rangle&=&
(\frac {u}{u-2})^{2L} (K_-(u)_{dd}+\frac{1}{u-1})|\Psi\rangle
-\frac{1}{u-1}|\Psi\rangle.
\eea
\bea
&&(u-1)T_{21}(u)\tilde{T}_{12}(u)-T_{22}(u)\tilde{T}_{22}(u)-
T_{23}(u)\tilde{T}_{32}(u)-
T_{24}(u)\tilde{T}_{42}(u) \no\\
&&~~~~~~~~~~~~~~~~~~~=-\tilde{T}_{11}(u)T_{11}(u)
+(u-1)\tilde{T}_{12}(u)T_{21}(u) -\tilde{T}_{13}(u)T_{31}(u)
-\tilde{T}_{14}(u)T_{41}(u),\no\\
&&(u-1)T_{21}(u)\tilde{T}_{13}(u)-T_{22}(u)\tilde{T}_{23}(u)-
T_{23}(u)\tilde{T}_{33}(u)-
T_{24}(u)\tilde{T}_{43}(u)=u\tilde{T}_{13}(u)T_{21}(u)\no\\
&&(u-1)T_{21}(u)\tilde{T}_{14}(u)-T_{22}(u)\tilde{T}_{24}(u)-
T_{23}(u)\tilde{T}_{34}(u)-
T_{24}(u)\tilde{T}_{44}(u)=u\tilde{T}_{14}(u)T_{21}(u)\no\\
&&T_{31}(u)\tilde{T}_{12}(u)-(u-1)T_{32}(u)\tilde{T}_{22}(u)+
T_{33}(u)\tilde{T}_{32}(u)+
T_{34}(u)\tilde{T}_{42}(u)=-u\tilde{T}_{22}(u)T_{32}(u)\no\\
&&T_{31}(u)\tilde{T}_{13}(u)-(u-1)T_{32}(u)\tilde{T}_{23}(u)+
T_{33}(u)\tilde{T}_{33}(u)+
T_{34}(u)\tilde{T}_{43}(u) \no\\
&&~~~~~~~~~~~~~~~~~~~=\tilde{T}_{21}(u)T_{12}(u)
+\tilde{T}_{22}(u)T_{22}(u) +(u+1)\tilde{T}_{23}(u)T_{32}(u)
+\tilde{T}_{24}(u)T_{42}(u),\no\\
&&T_{31}(u)\tilde{T}_{14}(u)-(u-1)T_{32}(u)\tilde{T}_{24}(u)+
T_{33}(u)\tilde{T}_{34}(u)+
T_{34}(u)\tilde{T}_{44}(u)=u\tilde{T}_{24}(u)T_{32}(u)\no\\
&&T_{41}(u)\tilde{T}_{12}(u)+T_{42}(u)\tilde{T}_{22}(u)+
(u+1)T_{43}(u)\tilde{T}_{32}(u)+
T_{44}(u)\tilde{T}_{42}(u)=u\tilde{T}_{32}(u)T_{43}(u)\no\\
&&T_{41}(u)\tilde{T}_{13}(u)+T_{42}(u)\tilde{T}_{23}(u)+
(u+1)T_{43}(u)\tilde{T}_{33}(u)+
T_{44}(u)\tilde{T}_{43}(u)=u\tilde{T}_{33}(u)T_{43}(u),\no\\
&&T_{41}(u)\tilde{T}_{14}(u)+T_{42}(u)\tilde{T}_{24}(u)+
(u+1)T_{43}(u)\tilde{T}_{34}(u)+
T_{44}(u)\tilde{T}_{44}(u) \no\\
&&~~~~~~~~~~~~~~~~~~~=\tilde{T}_{31}(u)T_{13}(u)
+\tilde{T}_{32}(u)T_{23}(u) +\tilde{T}_{33}(u)T_{33}(u)
+(u+1)\tilde{T}_{34}(u)T_{43}(u).
\eea
which come from a variant of the (graded) Yang-Baxter algebra
(\ref{rtt-ttr}) with the $R$ matrix (\ref {r1}),
\beq
\stackrel {1}{T}(u)R(2u)\stackrel {2}{\tilde{T}}(u)=
\stackrel {2}{\tilde{T}}(u)R(2u)\stackrel {1}{T}(u).
\eeq

Noticing the change $u \rightarrow u-1$ with respect to the
original problem,
one may check that these boundary $K$ matrices satisfy the reflection equations
for the reduced problem. After some algebra the reduced transfer matrix
$ \tau ^{(1)}(u)$ may be recognized as that for the inhomogeneous
supersymmetric $t-J$ open chain 
interacting with the Kondo impurities of arbitrary spins,
which has been diagonalized in Ref.\cite {zglg}.
The final result is
\bea
\Lambda ^{(1)}(u;\{ u_j \}) &=&
\frac {u}{u-2}\frac {(c_b-\frac {u}{2})}{(c_b-\frac
{u}{2}+2s_b)}
\frac {(c_b-\frac {u}{2}+2s_b+1)}{(c_b-\frac
{u}{2}-1)}
 \prod _{\a=1}^{M_1} \frac 
{(u-v_\a+2)(u+v_\a-2)}
{(u-v_\a)(u+v_\a-4)}\no\\
& &-\frac {u-1}{u-2} 
\prod ^N_{j=1} \frac {(u-u_j)(u+u_j-2)}
{(u-u_j-2)(u+u_j-4)}
\prod^{M_1} _{\a =1}
\frac {(u-v_\a+2)(u+v_\a-2)}
{(u-v_\a)(u+v_\a-4)}
\Lambda ^{(2)}(u;\{u_j\},\{v_\a\})
\eea
provided the parameters $\{ v_m \}$ satisfy 
\beq
\frac {v_\a}{v_\a-1} \frac {(c_b-\frac {v_\a}{2})(c_b -\frac {v_\a}{2}
+2s_b+1)}
{(c_b -\frac {v_\a}{2}+2s_b)(c_b-\frac {v_\a}{2}-1)}
\prod ^N_{j=1} \frac {(v_\a-u_j-2)(v_\a+u_j-4)}
{(v_\a-u_j)(v_\a+u_j-2)}=
-\Lambda ^{(2)}(v_\a;\{u_i\},\{v_\b\})
. \label {bethe2-1}
\eeq
Here $\Lambda ^{(2)}(u;\{ u_j \},\{ v_\a \})$ is the eigenvalue of the
transfer matrix $\tau ^{(2)}(u)$ for the $M_2$-site inhomogeneous $XXX$
open chain interacting with the Kondo impurities of arbitrary spins,
\bea
\Lambda ^{(2)}(u;\{ u_j \},\{v_\a \}) &=&
-\frac {(c_b-\frac {u}{2})}{(c_b-\frac
{u}{2}+2s_b)}
\frac {(c_b-\frac {u}{2}+2s_b+1)}{(c_b-\frac
{u}{2}-1)}\prod _{\g =a,b} \frac {c_\g+\frac {u}{2}+2s_\g-1}{c_\g-\frac
{u}{2}+2s_\g+1}\no\\
& &\{ \frac {u}{u-1} \prod _{\b=1}^{M_2} \frac 
{(u-w_\b-3)(u+w_\b-3)}
{(u-w_\b-1)(u+w_\b -1)}
+\frac {u-2}{u-1} \prod _{\g =a,b} \frac {(c_\g +\frac
{u}{2}-1)}{(c_\g-\frac {u}{2})}
 \frac {(c_\g -\frac
{u}{2}+2s_\g)}{(c_\g+\frac {u}{2}+2s_\g-1)}\no\\
& &\times\prod _{\a=1}^{M_1} \frac {(u-v_\a)(u+v_\a-4)}
{(u-v_\a+2)(u+v_\a-2)}
\prod ^{M_2}_{\b=1} \frac {(u-w_\b+1)(u+w_\b +1)}
{(u-w_\b -1)(u+w_\b -1)}\},
\eea
provided the parameters $\{w_\b \}$ satisfy 
\beq
\prod _{\g=a,b} \frac {(c_\g +\frac {w_\b}{2} -\frac {1}{2})
(c_\g-\frac {w_\b}{2}+2s_\g-\frac {1}{2})}
{(c_\g -\frac {w_\b}{2}-\frac {1}{2})
(c_\g +\frac {w_\b}{2}+2s_\g-\frac {1}{2})}
\prod ^{M_1}_{\a=1} \frac {(w_\b-v_\a +1)(w_\b +v_\a-3)}
 {(w_\b-v_\a +3)(w_\b +v_\a-1)}
=\prod ^{M_2}_{\stackrel {\d=1}{\d \neq \b}}
\frac {(w_\b -w_\d -2)
(w_\b +w_\d -2)}
{(w_\b -w_\d +2)
(w_\b +w_\d +2)}. \label {bethe3-1}
\eeq

After a shift of the parameters $u_j \rightarrow u_j+1,
v_m \rightarrow v_m + 2$, the Bethe ansatz equations (\ref
{bethe1-1}), (\ref{bethe2-1}) and 
(\ref {bethe3-1}) may be rewritten as follows
\bea
(\frac {u_j- 1}{u_j+1})^{2L}
\prod ^N_{\stackrel {i=1}{i \neq j}} \frac {(u_j-u_i +2)(u_j+u_i+2)}
{(u_j-u_i -2)(u_j+u_i-2)}&=& \prod ^{M_1}_{\a =1} \frac
{(u_j-v_\a+1)
(u_j+v_\a+1)}
{(u_j-v_\a-1)
(u_j+v_\a-1)},\no\\
\prod_{\g =a,b}
\frac{c_\g +\frac {v_\a}{2}+2s_\g}{c_\g-\frac{v_\a}{2}+2s_\g}
\prod ^N_{j=1} \frac {(v_\a -u_j+1)(v_\a+u_j+1)}
{(v_\a -u_j-1)(v_\a+u_j-1)}
& = & \prod_{\b=1}^{M_2} \frac{(v_\a -w_\b+1)(v_\a +w_\b +1)}
{(v_\a -w_\b-1)(v_\a +w_\b -1)},\no\\
\prod_{\g =a,b}
\frac{(c_\g+\frac {w_\b}{2}-\frac {1}{2})}
{(c_\g-\frac {w_\b}{2}-\frac {1}{2})}
\frac{(c_\g-\frac {w_\b}{2}+2s_\g-\frac {1}{2})}
{(c_\g+\frac {w_\b}{2}+2s_\g-\frac {1}{2})}
\prod_{\a=1}^{M_1} \frac{(w_\b -v_\a -1)}
{(w_\b -v_\a +1)}
\frac{(w_\b +v_\a -1)}
{(w_\b +v_\a +1)}
   &=&\prod ^{M_2} _{\stackrel {\d=1}{\d \neq \b}}
   \frac {(w_\b-w_\d-2)}{(w_\b -w_\d +2)}
   \frac {(w_\b+w_\d-2)}{(w_\b +w_\d +2)}
  ,\label{Bethe-ansatz1}
\eea
with the corresponding energy eigenvalue $E$ of the model 
\beq
E=-\sum ^N_{j=1} \frac {4}{u_j^2-1}.
\eeq

We now perform the algebraic Bethe ansatz method
\cite {Skl88,Gon94} procedure for the second couplings (\ref{ham2}).
Introducing the `doubled' monodromy matrix $U(u)$,
\beq
U(u)=T(u)K_-(u)\tilde{T}(u) \equiv
 \left ( \begin {array}
{cccc}
{\cal A}(u)&{\cal B}_{1}(u)&{\cal B}_2(u)&{\cal B}_3(u)\\
{\cal C}_{1}(u)&{\cal D}_{11}(u)&{\cal D}_{12}(u)&{\cal D}_{13}(u)\\
{\cal C}_{2}(u)&{\cal D}_{21}(u)&{\cal D}_{22}(u)&{\cal D}_{23}(u)\\
{\cal C}_{3}(u)&{\cal D}_{31}(u)&{\cal D}_{32}(u)&{\cal D}_{33}(u)\\
\end {array} \right ).
\eeq
where $\tilde {T}(u)=T^{-1}(-u)$.Substituting into the reflection equation
(\ref {reflection1}),we may draw the following
commutation relations,
\bea
{\check {\cal D}}_{bd}(u_1){\cal B}_c(u_2)&=&\frac {(u_1-u_2-2)(u_1+u_2-4)}
{(u_1-u_2)(u_1+u_2-2)}r(u_1+u_2-2)^{eb}_{gh}r(u_1-u_2)^{ih}_{cd}
{\cal B}_e(u_2){\check {\cal D}}_{gi}(u_1)-\no\\
& &\frac {2(u_1-2)u_2}{(u_1+u_2-2)(u_1-1)(u_2-1)}r(2u_1-2)^{gb}_{cd}
{\cal B}_g(u_1) {\cal A}(u_2) + \no\\
& & \frac {2(u_1-2)}{(u_1-u_2)(u_1-1)}
r(2u_1-2)^{gb}_{id} {\cal B}_g (u_1) {\check {\cal D}}_{ic}(u_2),\label
{cr2}\\
{\cal A}(u_1){\cal B}_{\b}(u_2) 
&=&\frac {(u_1-u_2+2)(u_1+u_2)} {(u_1-u_2)(u_1+u_2-2)}
{\cal B}_{\b}(u_2){\cal A}(u_1)-\frac {2(u_1+u_2)}{(u_1-u_2)(u_1+u_2-2)}
{\cal B}_{\b}(u_1)A(u_2)\no\\
& &  +\frac {2}{u_1+u_2-2}[{\cal B}_{\a}(u_1)({\check {\cal D}}_{\a \b}(u_2)
-\frac{1}{u_2-1}\d _{\a \b}{\cal A}(u_2)].
\eea
Here $ {\cal D} _{bd}(u) = {\check {\cal D}}_{bd}(u) - \frac {1}{u-1}
\delta _{bd}{\cal A}(u)$ and the matrix $r(u)$ ,which in turn satisfies the quantum Yang-Baxter
equation, takes the form,
\bea
r^{11}_{11}(u)&=&r^{22}_{22}(u)=r^{33}_{33}(u)=1,\no\\
 r^{12}_{12}(u)&=&r^{13}_{13}(u)=
r^{21}_{21}(u)=r^{31}_{31}(u)=
r^{23}_{23}(u)=r^{32}_{32}(u)=-\frac {2}{u-2},\no\\
r^{12}_{21}(u)&=&r^{21}_{12}(u)=
r^{13}_{31}(u)=r^{31}_{13}(u)=
r^{23}_{32}(u)=r^{32}_{23}(u)=
\frac {u}{u-2}.
\label{r4} 
\eea
 Choosing the Bethe state $|\Omega \rangle $ as
\beq
|\Omega \rangle = {\cal B}_{i_1}(u_1) \cdots {\cal
B}_{i_N}(u_N)|0\rangle F^{i_1\cdots i_N},
\eeq
with $|0\rangle $ being the pseudovacuum, and acting the transfer
matrix $\tau (u)$
on the state $|\Omega\rangle$,we have
$\tau (u) |\Omega \rangle =\Lambda(u) |\Omega \rangle$,with the
eigenvalue,
\bea
\Lambda (u)&=& \frac {u-4}{u-1}\frac {(c_b+\frac {u}{2}+s_b+\frac{1}{2})}
{(c_b+\frac {u}{2}+s_b-\frac{1}{2})}
\frac {(c_b+\frac {u}{2}-s_b-\frac{1}{2})}{(c_b+\frac
{u}{2}-s_b-\frac{3}{2})}
\prod ^N_{j=1} \frac {(u+u_j)(u-u_j+2)}{(u-u_j)(u+u_j-2)}\no\\
& &+\frac {u}{u-1} (\frac {u}{u-2})^{2L} 
\prod ^N_{j=1} \frac {(u-u_j-2)(u+u_j-4)}{(u-u_j)(u+u_j-2)}
\Lambda ^{(1)}(u;\{u_i\}),
\eea
provided the parameters $\{ u_j\}$ satisfy
\beq
\frac {u_j-4}{u_j-2}\frac {(c_b+\frac {u_j}{2}+s_b+\frac{1}{2})}
{(c_b+\frac {u_j}{2}+s_b-\frac{1}{2})}
\frac {(c_b+\frac {u_j}{2}-s_b-\frac{1}{2})}{(c_b+\frac
{u_j}{2}-s_b-\frac{3}{2})}
(\frac {u_j-2}{u_j})^{2L}=
\prod ^N_{\stackrel {i=1}{i \neq j}} \frac {(u_j-u_i-2)}{(u_j-u_i+2)}
\frac {(u_j+u_i-4)}{(u_j+u_i)}
\Lambda
^{(1)}(u_j;\{u_i\}). \label {bethe1-2}
\eeq
Here $\Lambda ^{(1)}(u;\{u_i\})$ is the eigenvalue of the transfer
matrix $\tau ^{(1)}(u)$ for the reduced problem,which arises out of the
$r$ matrices from the first term in the right hand side of (\ref {cr2}),with
the reduced boundary $K$ matrices $K_{\pm}^{(1)}(u)$ as,
\beq
K^{(1)}_-(u)=
  \left ( \begin {array}
{ccc}
1&0&0\\
0&A^{(1)}_-(u)&B^{(1)}_-(u)\\
0&C^{(1)}_-(u)&D^{(1)}_-(u)
\end {array} \right ),\label{k1-2}
\eeq
where
\bea
A^{(1)}_-(u)&=&-\frac {u^2-4c_a^{2}-4c_a+4s_a(s_a+1)+3-4(u-1) {\bf S}^z_a}
{(u+2c_a-2s_a-1)(u+2c_a+2s_a+1)},\no\\
B^{(1)}_-(u)&=&\frac {4(u-1) {\bf S}^-_a}
{(u+2c_a-2s_a-1)(u+2c_a+2s_a+1)},\no\\
C^{(1)}_-(u)&=&\frac {4(u-1) {\bf S}^+_a}
{(u+2c_a-2s_a-1)(u+2c_a+2s_a+1)},\no\\
D^{(1)}_-(u)&=&-\frac {u^2-4c_a^{2}-4c_a+4s_a(s_a+1)+3+4(u-1) {\bf S}^z_a}
{(u+2c_a-2s_a-1)(u+2c_a+2s_a+1)}.
\eea
and
\beq
K^{(1)}_+(u)=   \left ( \begin {array}
{ccc}
1&0&0\\
0&A^{(1)}_+(u)&B^{(1)}_+(u)\\
0&C^{(1)}_+(u)&D^{(1)}_+(u)
\end {array} \right ),
\eeq
where
\bea
A^{(1)}_+(u)&=&\frac{u^2-6u-4c^2_b-8c_b+4s_b(s_b+1)+5-4(u-4){\bf S}^z_b}
{(u+2c_b-2s_b-3)(u+2c_b+2s_b-1)},\no\\
B^{(1)}_+(u)&=&-\frac{4(u-4){\bf S}^-_b}
{(u+2c_b-2s_b-3)(u+2c_b+2s_b-1)},\no\\
C^{(1)}_+(u)&=&-\frac{4(u-4){\bf S}^+_b}
{(u+2c_b-2s_b-3)(u+2c_b+2s_b-1)},\no\\
D^{(1)}_+(u)&=&\frac{u^2-6u-4c^2_b-8c_b+4s_b(s_b+1)+5+4(u-4){\bf S}^z_b}
{(u+2c_b-2s_b-3)(u+2c_b+2s_b-1)}.
\eea
Here $K^{(1)}_- (u)$, the boundary $K$ matrices after  the first
nesting, follows from the relations,
\bea
{\check {\cal D}}_{dd}(u)|\Psi\rangle&\equiv&
\frac {u}{u-1}K^{(1)}_{dd}(u)|\Psi \rangle =
(K_-(u)_{dd}+\frac{1}{u-1})
(\frac{u}{u-2})^{2L}|\Psi\rangle,\no\\
{\check {\cal D}}_{db}(u)|\Psi\rangle&\equiv&
\frac {u}{u-1}K^{(1)}_{db}(u)|\Psi \rangle=
K_-(u)_{db} (\frac{u}{u-2})^{2L}|\Psi\rangle.
\eea
Indeed, applying the monodromy matrix
$T(u)$ and its ``adjoint'' ${\tilde T}(u)$ to the pseudovacuum, we have
\bea
T_{11}(u)|\Psi\rangle&=&|\Psi\rangle,~~~
T_{dd}(u)|\Psi\rangle=(\frac{u}{u-2})^L|\Psi\rangle,\no\\
T_{1d}(u)|\Psi\rangle&\neq& 0,~~~
T_{db}(u)|\Psi\rangle=0,~~~
T_{d1}(u)|\Psi\rangle= 0,\no\\
\tilde{T}_{11}(u)|\Psi\rangle&=&|\Psi\rangle,~~~
\tilde{T}_{dd}(u)|\Psi\rangle=(\frac{u}{u-2})^L|\Psi\rangle,\no\\
\tilde{T}_{1d}(u)|\Psi\rangle&\neq& 0,~~~
\tilde{T}_{db}(u)|\Psi\rangle=0,~~~
\tilde{T}_{d1}(u)|\Psi\rangle=0.
\eea
where $d\neq b,~~~ d,b=2,3,4$.Then we have
\bea
{\cal A}(u)|\Psi\rangle&=&|\Psi\rangle,\no\\
{\cal B}_d(u)|\Psi\rangle&\neq&0,~~~~~~
{\cal C}_d(u)|\Psi\rangle= 0,\no\\
{\cal D}_{db}(u)|\Psi\rangle&=&
(\frac{u}{u-2})^{2L} K_-(u)_{db}|\Psi\rangle,\no\\
{\cal D}_{dd}(u)|\Psi\rangle&=&
(\frac {u}{u-2})^{2L} (K_-(u)_{dd}+\frac{1}{u-1})|\Psi\rangle
-\frac{1}{u-1}|\Psi\rangle.
\eea
\bea
&&(u-1)T_{21}(u)\tilde{T}_{12}(u)-T_{22}(u)\tilde{T}_{22}(u)-
T_{23}(u)\tilde{T}_{32}(u)-
T_{24}(u)\tilde{T}_{42}(u) \no\\
&&~~~~~~~~~~~~~~~~~~~=-\tilde{T}_{11}(u)T_{11}(u)
+(u-1)\tilde{T}_{12}(u)T_{21}(u) -\tilde{T}_{13}(u)T_{31}(u)
-\tilde{T}_{14}(u)T_{41}(u),\no\\
&&(u-1)T_{21}(u)\tilde{T}_{13}(u)-T_{22}(u)\tilde{T}_{23}(u)-
T_{23}(u)\tilde{T}_{33}(u)-
T_{24}(u)\tilde{T}_{43}(u)=u\tilde{T}_{13}(u)T_{21}(u)\no\\
&&(u-1)T_{21}(u)\tilde{T}_{14}(u)-T_{22}(u)\tilde{T}_{24}(u)-
T_{23}(u)\tilde{T}_{34}(u)-
T_{24}(u)\tilde{T}_{44}(u)=u\tilde{T}_{14}(u)T_{21}(u)\no\\
&&T_{31}(u)\tilde{T}_{12}(u)-(u-1)T_{32}(u)\tilde{T}_{22}(u)+
T_{33}(u)\tilde{T}_{32}(u)+
T_{34}(u)\tilde{T}_{42}(u)=-u\tilde{T}_{22}(u)T_{32}(u)\no\\
&&T_{31}(u)\tilde{T}_{13}(u)-(u-1)T_{32}(u)\tilde{T}_{23}(u)+
T_{33}(u)\tilde{T}_{33}(u)+
T_{34}(u)\tilde{T}_{43}(u) \no\\
&&~~~~~~~~~~~~~~~~~~~=\tilde{T}_{21}(u)T_{12}(u)
+\tilde{T}_{22}(u)T_{22}(u) -(u-1)\tilde{T}_{23}(u)T_{32}(u)
+\tilde{T}_{24}(u)T_{42}(u),\no\\
&&T_{31}(u)\tilde{T}_{14}(u)-(u-1)T_{32}(u)\tilde{T}_{24}(u)+
T_{33}(u)\tilde{T}_{34}(u)+
T_{34}(u)\tilde{T}_{44}(u)=-u\tilde{T}_{24}(u)T_{32}(u)\no\\
&&T_{41}(u)\tilde{T}_{12}(u)+T_{42}(u)\tilde{T}_{22}(u)
-(u-1)T_{43}(u)\tilde{T}_{32}(u)+
T_{44}(u)\tilde{T}_{42}(u)=u\tilde{T}_{32}(u)T_{43}(u)\no\\
&&T_{41}(u)\tilde{T}_{13}(u)+T_{42}(u)\tilde{T}_{23}(u)
-(u-1)T_{43}(u)\tilde{T}_{33}(u)+
T_{44}(u)\tilde{T}_{43}(u)=-u\tilde{T}_{33}(u)T_{43}(u),\no\\
&&T_{41}(u)\tilde{T}_{14}(u)+T_{42}(u)\tilde{T}_{24}(u)
-(u-1)T_{43}(u)\tilde{T}_{34}(u)+
T_{44}(u)\tilde{T}_{44}(u) \no\\
&&~~~~~~~~~~~~~~~~~~~=\tilde{T}_{31}(u)T_{13}(u)
+\tilde{T}_{32}(u)T_{23}(u) +\tilde{T}_{33}(u)T_{33}(u)
-(u-1)\tilde{T}_{34}(u)T_{43}(u).
\eea
which come from a variant of the (graded) Yang-Baxter algebra
(\ref{rtt-ttr}) with the $R$ matrix (\ref {r2}),
\beq
\stackrel {1}{T}(u)R(2u)\stackrel {2}{\tilde{T}}(u)=
\stackrel {2}{\tilde{T}}(u)R(2u)\stackrel {1}{T}(u).
\eeq

Noticing the change $u \rightarrow u-1$ with respect to the
original problem,
one may check that these boundary $K$ matrices satisfy the reflection equations
for the reduced problem. After some algebra,the reduced transfer matrix
$ \tau ^{(1)}(u)$ may be recognized as that for the inhomogeneous
$su(3) \; t-J$ open chain 
interacting with the Kondo impurities of arbitrary spins,
which has been diagonalized in Ref.\cite {zglg}.
The final result is,
\bea
\Lambda ^{(1)}(u;\{ u_j \}) &=&
\frac {u-4}{u-2}\frac {(c_b+\frac {u}{2}+s_b+\frac{1}{2})}{(c_b+\frac
{u}{2}+s_b-\frac{1}{2})}
\frac {(c_b+\frac {u}{2}-s_b-\frac{1}{2})}{(c_b+\frac
{u}{2}-s_b-\frac{3}{2})}
 \prod _{\a=1}^{M_1} \frac 
{(u-v_\a+2)(u+v_\a-2)}
{(u-v_\a)(u+v_\a-4)}\no\\
& &-\frac {u-1}{u-2} 
\prod ^N_{j=1} \frac {(u-u_j)(u+u_j-2)}
{(u-u_j-2)(u+u_j-4)}
\prod^{M_1} _{\a =1}
\frac {(u-v_\a-2)(u+v_\a-6)}
{(u-v_\a)(u+v_\a-4)}
\Lambda ^{(2)}(u;\{u_j\},\{v_\a\})
\eea
provided the parameters $\{ v_m \}$ satisfy 
\bea
\frac {v_\a-4}{v_\a-3}\frac {(c_b+\frac {v_\a}{2}+s_b+\frac{1}{2})}{(c_b+\frac
{v_\a}{2}+s_b-\frac{1}{2})}
\frac {(c_b+\frac {v_\a}{2}-s_b-\frac{1}{2})}{(c_b+\frac
{v_\a}{2}-s_b-\frac{3}{2})}
& &\prod ^N_{j=1} \frac {(v_\a-u_j-2)(v_\a+u_j-4)}
{(v_\a-u_j)(v_\a+u_j-2)}
\prod ^{M_1}_{\stackrel {\z=1}{\z \neq \a}}
\frac {(v_\a -v_\z +2)
(v_\a +v_\z -2)}
{(v_\a -v_\z -2)
(v_\a +v_\z-6)}\no\\
&=&-\Lambda ^{(2)}(v_\a;\{u_i\},\{v_\b\})
. \label {bethe2-2}
\eea
Here $\Lambda ^{(2)}(u;\{ u_j \},\{ v_\a \})$ is the eigenvalue of the
transfer matrix $\tau ^{(2)}(u)$ for the $M_2$-site inhomogeneous $XXX$
open chain
interacting with the Kondo impurities of arbitrary spins,
\bea
\Lambda ^{(2)}(u;\{ u_j \},\{v_\a \}) &=&
-\frac {(c_b+\frac {u}{2}+s_b+\frac{1}{2})}{(c_b+\frac
{u}{2}+s_b-\frac{1}{2})}
\frac {(c_b+\frac {u}{2}-s_b-\frac{1}{2})}{(c_b+\frac
{u}{2}-s_b-\frac{3}{2})}
\prod _{\g =a,b} \frac {c_\g-\frac {u}{2}+s_\g +\frac{5}{2}}{c_\g+\frac
{u}{2}+s_\g +\frac{1}{2}}\no\\
& &\{ \frac {u-4}{u-3} \prod _{\b=1}^{M_2} \frac 
{(u-w_\b+2)(u+w_\b-4)}
{(u-w_\b)(u+w_\b -6)}
+\frac {u-2}{u-3} \prod _{\g =a,b} \frac {(c_\g +\frac
{u}{2}+s_\g -\frac{1}{2})}{(c_\g+\frac {u}{2}-s_\g -\frac{1}{2})}
 \frac {(c_\g -\frac
{u}{2}-s_\g+\frac{5}{2})}{(c_\g-\frac {u}{2}+s_\g+\frac{5}{2})}\no\\
& &\times\prod _{\a=1}^{M_1} \frac {(u-v_\a)(u+v_\a-4)}
{(u-v_\a-2)(u+v_\a-6)}
\prod ^{M_2}_{\b=1} \frac {(u-w_\b-2)(u+w_\b -8)}
{(u-w_\b)(u+w_\b -6)}\},
\eea
provided the parameters $\{w_\b \}$ satisfy 
\beq
\prod _{\g=a,b} \frac {(c_\g +\frac {w_\b}{2} +s_\g-\frac {1}{2})
(c_\g-\frac {w_\b}{2}-s_\g+\frac {5}{2})}
{(c_\g -\frac {w_\b}{2}+s_\g+\frac {5}{2})
(c_\g +\frac {w_\b}{2}-s_\g-\frac {1}{2})}
\prod ^{M_1}_{\a=1} \frac {(w_\b-v_\a )(w_\b +v_\a-4)}
 {(w_\b-v_\a -2)(w_\b +v_\a-6)}
=\prod ^{M_2}_{\stackrel {\d=1}{\d \neq \b}}
\frac {(w_\b -w_\d +2)
(w_\b +w_\d -4)}
{(w_\b -w_\d-2 )
(w_\b +w_\d -8)}. \label {bethe3-2}
\eeq

After a shift of the parameters $u_j \rightarrow u_j+1,
v_m \rightarrow v_m + 2,
w_l \rightarrow w_l + 3$, the Bethe ansatz equations (\ref
{bethe1-2}) , 
(\ref {bethe2-2}) and (\ref {bethe3-2}) may be rewritten as follows
\bea
(\frac {u_j- 1}{u_j+1})^{2L}
\prod ^N_{\stackrel {i=1}{i \neq j}} \frac {(u_j-u_i +2)(u_j+u_i+2)}
{(u_j-u_i -2)(u_j+u_i-2)}&=& \prod ^{M_1}_{\a =1} \frac
{(u_j-v_\a+1)
(u_j+v_\a+1)}
{(u_j-v_\a-1)
(u_j+v_\a-1)},\no\\
\prod_{\g =a,b}
\frac{c_\g +\frac {v_\a}{2}+s_\g +\frac{3}{2}}{c_\g-\frac{v_\a}{2}+s_\g+
\frac{3}{2}}
\prod ^N_{j=1} \frac {(v_\a -u_j-1)(v_\a+u_j-1)}
{(v_\a -u_j+1)(v_\a+u_j+1)}
& = & \prod_{\b=1}^{M_2} \frac{(v_\a -w_\b+1)(v_\a +w_\b +1)}
{(v_\a -w_\b-1)(v_\a +w_\b -1)},\no\\
& &\times\prod ^{M_1}_{\stackrel {\z=1}{\z \neq \a}}
\frac {(v_\a -v_\z -2)
(v_\a +v_\z -2)}
{(v_\a -w_\z +2)
(v_\a +v_\z+2)}\no\\
\prod_{\g =a,b}\frac{(c_\g+\frac {w_\b}{2}+s_\g+1)
(c_\g-\frac {w_\b}{2}-s_\g+1)}
{(c_\g-\frac {w_\b}{2}+s_\g+1)
(c_\g+\frac {w_\b}{2}-s_\g+1)}
\prod_{\a=1}^{M_1} \frac{(w_\b -v_\a +1)}
{(w_\b -v_\a -1)}
\frac{(w_\b +v_\a +1)}
{(w_\b +v_\a- 1)}
   &=&\prod ^{M_2}_{\stackrel {\d=1}{\d \neq \b}}
   \frac {(w_\b-w_\d+2)}{(w_\b -w_\d -2)}
   \frac {(w_\b+w_\d+2)}{(w_\b +w_\d -2)}
  ,\label{Bethe-ansatz2}
\eea
with the corresponding energy eigenvalue $E$ of the model 
\beq
E=-\sum ^N_{j=1} \frac {4}{u_j^2-1}.
\eeq

We now perform the algebraic Bethe ansatz method
\cite {Skl88,Gon94} procedure for the third couplings (\ref{ham3}).
Introducing the `doubled' monodromy matrix $U(u)$,
\beq
U(u)=T(u)K_-(u)\tilde{T}(u) \equiv
 \left ( \begin {array}
{cccc}
{\cal A}(u)&{\cal B}_{1}(u)&{\cal B}_2(u)&{\cal B}_3(u)\\
{\cal C}_{1}(u)&{\cal D}_{11}(u)&{\cal D}_{12}(u)&{\cal D}_{13}(u)\\
{\cal C}_{2}(u)&{\cal D}_{21}(u)&{\cal D}_{22}(u)&{\cal D}_{23}(u)\\
{\cal C}_{3}(u)&{\cal D}_{31}(u)&{\cal D}_{32}(u)&{\cal D}_{33}(u)\\
\end {array} \right ).
\eeq
where $\tilde {T}(u)=T^{-1}(-u)$.Substituting into the reflection equation
(\ref {reflection1}),we may draw the following
commutation relations,
\bea
{\check {\cal D}}_{bd}(u_1){\cal B}_c(u_2)&=&\frac {(u_1-u_2-2)(u_1+u_2)}
{(u_1-u_2)(u_1+u_2+2)}r(u_1+u_2+2)^{eb}_{gh}r(u_1-u_2)^{ih}_{cd}
{\cal B}_e(u_2){\check {\cal D}}_{gi}(u_1)-\no\\
& &\frac {2u_1u_2}{(u_1+u_2+2)(u_1+1)(u_2+1)}r(2u_1+2)^{gb}_{cd}
{\cal B}_g(u_1) {\cal A}(u_2) + \no\\
& & \frac {2u_1}{(u_1-u_2)(u_1+1)}
r(2u_1+2)^{gb}_{id} {\cal B}_g (u_1) {\check {\cal D}}_{ic}(u_2),\label
{cr}\\
{\cal A}(u_1){\cal B}_{\b}(u_2) 
&=&\frac {(u_1-u_2-2)(u_1+u_2)} {(u_1-u_2)(u_1+u_2+2)}
{\cal B}_{\b}(u_2){\cal A}(u_1)-\frac {2(u_1+u_2)}{(u_1-u_2)(u_1+u_2+2)}
{\cal B}_{\b}(u_1)A(u_2)\no\\
& &  -\frac {2}{u_1+u_2+2}[{\cal B}_{\a}(u_1)({\check {\cal D}}_{\a \b}(u_2)
-\frac{1}{u_2-1}\d _{\a \b}{\cal A}(u_2)].
\eea
Here $ {\cal D} _{bd}(u) = {\check {\cal D}}_{bd}(u) + \frac {1}{u+1}
\delta _{bd}{\cal A}(u)$ and the matrix $r(u)$ ,which in turn satisfies
the quantum Yang-Baxter equation, takes the form,
\bea
r^{11}_{11}(u)&=&r^{22}_{22}(u)=r^{33}_{33}(u)=1,\no\\
 r^{12}_{12}(u)&=&r^{13}_{13}(u)=
r^{21}_{21}(u)=r^{31}_{31}(u)=
r^{23}_{23}(u)=r^{32}_{32}(u)=-\frac {2}{u-2},\no\\
r^{12}_{21}(u)&=&r^{21}_{12}(u)=
r^{13}_{31}(u)=r^{31}_{13}(u)=
r^{23}_{32}(u)=r^{32}_{23}(u)=
\frac {u}{u-2}.
\eea
 Choosing the Bethe state $|\Omega \rangle $ as
\beq
|\Omega \rangle = {\cal B}_{i_1}(u_1) \cdots {\cal
B}_{i_N}(u_N)|0\rangle F^{i_1\cdots i_N},
\eeq
with $|0\rangle $ being the pseudovacuum, and acting the transfer
matrix $\tau (u)$
on the state $|\Omega\rangle$,we have
$\tau (u) |\Omega \rangle =\Lambda(u) |\Omega \rangle$,with the
eigenvalue,
\bea
\Lambda (u)&=& \frac {u-2}{u+1}\frac {(c_b+\frac {u}{2}-s_b-\frac{1}{2})}
{(c_b+\frac {u}{2}+s_b-\frac{1}{2})}
\frac {(c_b+\frac {u}{2}+s_b+\frac{1}{2})}{(c_b+\frac
{u}{2}-s_b-\frac{3}{2})}
\prod ^N_{j=1} \frac {(u+u_j)(u-u_j-2)}{(u-u_j)(u+u_j+2)}\no\\
& &+\frac {u}{u+1} (-\frac {u}{u+2})^{2L} 
\prod ^N_{j=1} \frac {(u+u_j)(u-u_j-2)}{(u-u_j)(u+u_j+2)}
\Lambda ^{(1)}(u;\{u_i\}),
\eea
provided the parameters $\{ u_j\}$ satisfy
\beq
 \frac {u_j-2}{u_j}\frac {(c_b+\frac {u_j}{2}-s_b-\frac{1}{2})}
{(c_b+\frac {u_j}{2}+s_b-\frac{1}{2})}
\frac {(c_b+\frac {u_j}{2}+s_b+\frac{1}{2})}{(c_b+\frac
{u_j}{2}-s_b-\frac{3}{2})}
(-\frac {u_j+2}{u_j})^{2L}=
\prod ^N_{\stackrel {i=1}{i \neq j}} \frac {(u_j-u_i+2)}{(u_j-u_i-2)}
\frac {(u_j+u_i+4)}{(u_j+u_i)}
\Lambda
^{(1)}(u_j;\{u_i\}). \label {bethe1-3}
\eeq
Here $\Lambda ^{(1)}(u;\{u_i\})$ is the eigenvalue of the transfer
matrix $\tau ^{(1)}(u)$ for the reduced problem,which arises out of the
$r$ matrices from the first term in the right hand side of (\ref {cr}),with
the reduced boundary $K$ matrices $K_{\pm}^{(1)}(u)$ as,
\beq
K^{(1)}_-(u)=
  \left ( \begin {array}
{ccc}
1&0&0\\
0&A^{(1)}_-(u)&B^{(1)}_-(u)\\
0&C^{(1)}_-(u)&D^{(1)}_-(u)
\end {array} \right ),\label{k1-3}
\eeq
where
\bea
A^{(1)}_-(u)&=&-\frac {u^2-4c_a^{2}+4c_a+4s_a(s_a+1)-1-4(u+1) {\bf S}^z_a}
{(u+2c_a-2s_a-1)(u+2c_a+2s_a+1)},\no\\
B^{(1)}_-(u)&=&\frac {4(u+1) {\bf S}^-_a}
{(u+2c_a-2s_a-1)(u+2c_a+2s_a+1)},\no\\
C^{(1)}_-(u)&=&\frac {4(u+1) {\bf S}^+_a}
{(u+2c_a-2s_a-1)(u+2c_a+2s_a+1)},\no\\
D^{(1)}_-(u)&=&-\frac {u^2-4c_a^{2}+4c_a+4s_a(s_a+1)-1+4(u+1) {\bf S}^z_a}
{(u+2c_a-2s_a-1)(u+2c_a+2s_a+1)}.
\eea
and
\beq
K^{(1)}_+(u)=   \left ( \begin {array}
{ccc}
-1&0&0\\
0&A^{(1)}_+(u)&B^{(1)}_+(u)\\
0&C^{(1)}_+(u)&D^{(1)}_+(u)
\end {array} \right ),
\eeq
where
\bea
A^{(1)}_+(u)&=&-\frac{u^2-2u-4c^2_b+4s_b(s_b+1)+1-4(u-2){\bf S}^z_b}
{(u+2c_b-2s_b-3)(u+2c_b+2s_b-1)},\no\\
B^{(1)}_+(u)&=&\frac{4(u-2){\bf S}^-_b}
{(u+2c_b-2s_b-3)(u+2c_b+2s_b-1)},\no\\
C^{(1)}_+(u)&=&\frac{4(u-2){\bf S}^+_b}
{(u+2c_b-2s_b-3)(u+2c_b+2s_b-1)},\no\\
D^{(1)}_+(u)&=&-\frac{u^2-2u-4c^2_b+4s_b(s_b+1)+1+4(u-2){\bf S}^z_b}
{(u+2c_b-2s_b-3)(u+2c_b+2s_b-1)}.
\eea
Here $K^{(1)}_- (u)$, the boundary $K$ matrices after  the first
nesting, follows from the relations,
\bea
{\check {\cal D}}_{dd}(u)|\Psi\rangle&\equiv&
\frac {u}{u+1}K^{(1)}_{dd}(u)|\Psi \rangle =
(K_-(u)_{dd}-\frac{1}{u+1})
(-\frac{u}{u+2})^{2L}|\Psi\rangle,\no\\
{\check {\cal D}}_{db}(u)|\Psi\rangle&\equiv&
\frac {u}{u+1}K^{(1)}_{db}(u)|\Psi \rangle=
K_-(u)_{db} (-\frac{u}{u+2})^{2L}|\Psi\rangle.
\eea
Indeed, applying the monodromy matrix
$T(u)$ and its ``adjoint'' ${\tilde T}(u)$ to the pseudovacuum, we have
\bea
T_{11}(u)|\Psi\rangle&=&|\Psi\rangle,~~~
T_{dd}(u)|\Psi\rangle=(-\frac{u}{u+2})^L|\Psi\rangle,\no\\
T_{1d}(u)|\Psi\rangle&\neq& 0,~~~
T_{db}(u)|\Psi\rangle=0,~~~
T_{d1}(u)|\Psi\rangle= 0,\no\\
\tilde{T}_{11}(u)|\Psi\rangle&=&|\Psi\rangle,~~~
\tilde{T}_{dd}(u)|\Psi\rangle=(-\frac{u}{u+2})^L|\Psi\rangle,\no\\
\tilde{T}_{1d}(u)|\Psi\rangle&\neq& 0,~~~
\tilde{T}_{db}(u)|\Psi\rangle=0,~~~
\tilde{T}_{d1}(u)|\Psi\rangle=0.
\eea
where $d\neq b,~~~ d,b=2,3,4$.Then we have
\bea
{\cal A}(u)|\Psi\rangle&=&|\Psi\rangle,\no\\
{\cal B}_d(u)|\Psi\rangle&\neq&0,~~~~~~
{\cal C}_d(u)|\Psi\rangle= 0,\no\\
{\cal D}_{db}(u)|\Psi\rangle&=&
(-\frac{u}{u+2})^{2L} K_-(u)_{db}|\Psi\rangle,\no\\
{\cal D}_{dd}(u)|\Psi\rangle&=&
(-\frac {u}{u+2})^{2L} (K_-(u)_{dd}-\frac{1}{u+1})|\Psi\rangle
+\frac{1}{u+1}|\Psi\rangle.
\eea
\bea
&&(u+1)T_{21}(u)\tilde{T}_{12}(u)+T_{22}(u)\tilde{T}_{22}(u)+
T_{23}(u)\tilde{T}_{32}(u)+
T_{24}(u)\tilde{T}_{42}(u) \no\\
&&~~~~~~~~~~~~~~~~~~~=\tilde{T}_{11}(u)T_{11}(u)
-(u-1)\tilde{T}_{12}(u)T_{21}(u) +\tilde{T}_{13}(u)T_{31}(u)
+\tilde{T}_{14}(u)T_{41}(u),\no\\
&&(u+1)T_{21}(u)\tilde{T}_{13}(u)+T_{22}(u)\tilde{T}_{23}(u)+
T_{23}(u)\tilde{T}_{33}(u)+
T_{24}(u)\tilde{T}_{43}(u)=-u\tilde{T}_{13}(u)T_{21}(u)\no\\
&&(u+1)T_{21}(u)\tilde{T}_{14}(u)+T_{22}(u)\tilde{T}_{24}(u)+
T_{23}(u)\tilde{T}_{34}(u)+
T_{24}(u)\tilde{T}_{44}(u)=-u\tilde{T}_{14}(u)T_{21}(u)\no\\
&&T_{31}(u)\tilde{T}_{12}(u)-(u-1)T_{32}(u)\tilde{T}_{22}(u)+
T_{33}(u)\tilde{T}_{32}(u)+
T_{34}(u)\tilde{T}_{42}(u)=-u\tilde{T}_{22}(u)T_{32}(u)\no\\
&&T_{31}(u)\tilde{T}_{13}(u)-(u-1)T_{32}(u)\tilde{T}_{23}(u)+
T_{33}(u)\tilde{T}_{33}(u)+
T_{34}(u)\tilde{T}_{43}(u) \no\\
&&~~~~~~~~~~~~~~~~~~~=\tilde{T}_{21}(u)T_{12}(u)
+\tilde{T}_{22}(u)T_{22}(u) -(u-1)\tilde{T}_{23}(u)T_{32}(u)
+\tilde{T}_{24}(u)T_{42}(u),\no\\
&&T_{31}(u)\tilde{T}_{14}(u)-(u-1)T_{32}(u)\tilde{T}_{24}(u)+
T_{33}(u)\tilde{T}_{34}(u)+
T_{34}(u)\tilde{T}_{44}(u)=-u\tilde{T}_{24}(u)T_{32}(u)\no\\
&&T_{41}(u)\tilde{T}_{12}(u)+T_{42}(u)\tilde{T}_{22}(u)-
(u-1)T_{43}(u)\tilde{T}_{32}(u)+
T_{44}(u)\tilde{T}_{42}(u)=u\tilde{T}_{32}(u)T_{43}(u)\no\\
&&T_{41}(u)\tilde{T}_{13}(u)+T_{42}(u)\tilde{T}_{23}(u)-
(u-1)T_{43}(u)\tilde{T}_{33}(u)+
T_{44}(u)\tilde{T}_{43}(u)=-u\tilde{T}_{33}(u)T_{43}(u),\no\\
&&T_{41}(u)\tilde{T}_{14}(u)+T_{42}(u)\tilde{T}_{24}(u)
-(u-1)T_{43}(u)\tilde{T}_{34}(u)+
T_{44}(u)\tilde{T}_{44}(u) \no\\
&&~~~~~~~~~~~~~~~~~~~=\tilde{T}_{31}(u)T_{13}(u)
+\tilde{T}_{32}(u)T_{23}(u) +\tilde{T}_{33}(u)T_{33}(u)
-(u-1)\tilde{T}_{34}(u)T_{43}(u).
\eea
which come from a variant of the (graded) Yang-Baxter algebra
(\ref{rtt-ttr}) with the $R$ matrix (\ref {r3}),
\beq
\stackrel {1}{T}(u)R(2u)\stackrel {2}{\tilde{T}}(u)=
\stackrel {2}{\tilde{T}}(u)R(2u)\stackrel {1}{T}(u).
\eeq

Noticing the change $u \rightarrow u+1$ with respect to the
original problem,
one may check that these boundary $K$ matrices satisfy the reflection equations
for the reduced problem. After some algebra,the reduced transfer matrix
$ \tau ^{(1)}(u)$ may be recognized as that for the inhomogeneous
$su(3)\; t-J$ open chain 
interacting with the Kondo impurities of arbitrary spins,
which has been diagonalized in Ref.\cite {zglg}.
The final result is,
\bea
\Lambda ^{(1)}(u;\{ u_j \}) &=&
\frac {u-2}{u}\frac {(c_b+\frac {u}{2}-s_b-\frac{1}{2})}{(c_b+\frac
{u}{2}+s_b-\frac{1}{2})}
\frac {(c_b+\frac {u}{2}+s_b+\frac{1}{2})}{(c_b+\frac
{u}{2}-s_b-\frac{3}{2})}
 \prod _{\a=1}^{M_1} \frac 
{(u-v_\a+2)(u+v_\a+2)}
{(u-v_\a)(u+v_\a)}\no\\
& &-\frac {u+1}{u} 
\prod ^N_{j=1} \frac {(u-u_j)(u+u_j+2)}
{(u-u_j-2)(u+u_j)}
\prod^{M_1} _{\a =1}
\frac {(u-v_\a-2)(u+v_\a-2)}
{(u-v_\a)(u+v_\a)}
\Lambda ^{(2)}(u;\{u_j\},\{v_\a\})
\eea
provided the parameters $\{ v_m \}$ satisfy 
\bea
\frac {v_\a-2}{v_\a-1}\frac {(c_b+\frac {v_\a}{2}-s_b-\frac{1}{2})}{(c_b+\frac
{v_\a}{2}+s_b-\frac{1}{2})}
\frac {(c_b+\frac {v_\a}{2}+s_b+\frac{1}{2})}{(c_b+\frac
{v_\a}{2}-s_b-\frac{3}{2})}
& &\prod ^N_{j=1} \frac {(v_\a-u_j-2)(v_\a+u_j)}
{(v_\a-u_j)(v_\a+u_j+2)}
\prod ^{M_1}_{\stackrel {\z=1}{\z \neq \a}}
\frac {(v_\a-v_\z+2)
(v_\a+v_\z+2)}
{(v_\a -v_\z-2)
(v_\a+v_\z-2)}\no\\
&=& -\Lambda ^{(2)}(v_\a;\{u_i\},\{v_\b\})
. \label {bethe2-3}
\eea
Here $\Lambda ^{(2)}(u;\{ u_j \},\{ v_\a \})$ is the eigenvalue of the
transfer matrix $\tau ^{(2)}(u)$ for the $M_2$-site inhomogeneous $XXX$
open chain
interacting with the Kondo impurities of arbitrary spins,
\bea
\Lambda ^{(2)}(u;\{ u_j \},\{v_\a \}) &=&
-\frac {(c_b+\frac {u}{2}-s_b-\frac{1}{2})}{(c_b+\frac
{u}{2}+s_b-\frac{1}{2})}
\frac {(c_b+\frac {u}{2}+s_b+\frac{1}{2})}{(c_b+\frac
{u}{2}-s_b-\frac{3}{2})}
\prod _{\g =a,b} \frac {c_\g-\frac {u}{2}+s_\g+\frac{1}{2}}{c_\g+\frac
{u}{2}+s_\g+\frac{1}{2}} \no\\
& &\{\frac {u-2}{u-1} \prod ^{M_2}_{\b=1} \frac 
{(u-w_\b+2)(u+w_\b)}
{(u-w_\b)(u+w_\b-2)}
+\frac {u}{u-1} \prod _{\g =a,b} \frac {(c_\g +\frac
{u}{2}+s_\g-\frac{1}{2})}{(c_\g+\frac {u}{2}-s_\g-\frac{1}{2})}
 \frac {(c_\g -\frac {u}{2}-s_\g+\frac{1}{2})}{(c_\g-\frac {u}{2}+s_\g+
 \frac{1}{2})}\no\\
& &\times\prod _{\a=1}^{M_1} \frac {(u-v_\a)(u+v_\a)}
{(u-v_\a-2)(u+v_\a-2)}
\prod ^{M_2}_{\b=1} \frac {(u-w_\b-2)(u+w_\b -4)}
{(u-w_\b )(u+w_\b -2)}\},
\eea
provided the parameters $\{w_\b \}$ satisfy 
\beq
\prod _{\g=a,b} \frac {(c_\g +\frac {w_\b}{2}+s_\g -\frac {1}{2})
(c_\g-\frac {w_\b}{2}-s_\g+\frac {1}{2})}
{(c_\g -\frac {w_\b}{2}+s_\g+\frac {1}{2})
(c_\g+\frac {w_\b}{2}-s_\g-\frac {1}{2})}
\prod ^{M_1}_{\a=1} \frac {(w_\b-v_\a)(w_\b +v_\a)}
 {(w_\b-v_\a -2)(w_\b +v_\a-2)}
=\prod ^{M_2}_{\stackrel {\d=1}{\d \neq \b}}
\frac {(w_\b -w_\d +2)
(w_\b +w_\d )}
{(w_\b -w_\d -2)
(w_\b +w_\d -4)}. \label {bethe3-3}
\eeq

After a shift of the parameters $u_j \rightarrow u_j-1,
w_m \rightarrow w_m +1$, the Bethe ansatz equations (\ref
{bethe1-3}), 
(\ref {bethe2-3}) and (\ref{bethe3-3}) may be rewritten as follows
\bea
(\frac {u_j+ 1}{u_j-1})^{2L}
\prod ^N_{\stackrel {i=1}{i \neq j}} \frac {(u_j-u_i +2)(u_j+u_i+2)}
{(u_j-u_i -2)(u_j+u_i-2)}&=& \prod ^{M_1}_{\a =1} \frac
{(u_j-v_\a+1)
(u_j+v_\a+1)}
{(u_j+v_\a-1)
(u_j-v_\a-1)},\no\\
\prod_{\g =a,b}
\frac{c_\g +\frac {v_\a}{2}+s_\g+\frac{1}{2}}{c_\g-\frac{v_\a}{2}+s_\g+
\frac{1}{2}}
\prod ^N_{j=1} \frac {(v_\a -u_j-1)(v_\a+u_j-1)}
{(v_\a -u_j+1)(v_\a+u_j+1)}
& = & \prod_{\b=1}^{M_2} \frac{(v_\a -w_\b+1)(v_\a +w_\b +1)}
{(v_\a -w_\b-1)(v_\a +w_\b -1)},\no\\
& &\times\prod ^{M_1}_{\stackrel {\z=1}{\z \neq \a}}
\frac {(v_\a-v_\z-2) (v_\a+v_\z-2 )}
{(v_\a-v_\z+2) (v_\a +v_\z+2)}\no\\
\prod_{\g =a,b}
\frac{(c_\g+\frac {w_\b}{2}+s_\g)}
{(c_\g-\frac {w_\b}{2}+s_\g)}
\frac{(c_\g-\frac {w_\b}{2}-s_\g)}
{(c_\g+\frac {w_\b}{2}-s_\g)}
\prod_{\a=1}^{M_1} \frac{(w_\b -v_\a +1)}
{(w_\b -v_\a -1)}
\frac{(w_\b +v_\a +1)}
{(w_\b +v_\a -1)}
   &=&\prod ^{M_2}_{\stackrel {\d=1}{\d \neq \b}}
   \frac {(w_\b-w_\d+2)}{(w_\b -w_\d -2)}
   \frac {(w_\b+w_\d+2)}{(w_\b +w_\d -2)}
  ,\label{Bethe-ansatz3}
\eea
with the corresponding energy eigenvalue $E$ of the model 
\beq
E=-\sum ^N_{j=1} \frac {4}{u_j^2-1}.
\eeq

\section{Conclusion}

In conclusion, we have studied integrable Kondo problems describing two
boundary impurities coupled to one-dimensional extended Hubbard open chains. 
The  quantum integrability of these
systems follows from the fact
that the Hamiltonians in each case are derived from  
a one-parameter family of commuting transfer matrices. Moreover, the Bethe
Ansatz equations and expressions for the energies 
are derived by means of the algebraic Bethe ansatz
approach. We would like to  emphasize that the boundary $K$ matices found here
are non-regular in that they can not be factorized into the product
of a c-number $K$ matrix and the local momodromy matrices. However, 
similar to the cases discussed in \cite{ZG,zglg}, it is
 possible to introduce a singular local monodromy matrix $\tilde
L(u)$
to express the boundary $K$
matrix $K_-(u)$ as,
\beq
K_-(u)=\tilde {L}(u){\tilde {L}}^{-1}(-u).
\eeq
where, for example in the case of the superalgebras $gl(2|2)$ model   
\beq
\tilde L (u) =
 \left ( \begin {array}
{cccc}
\e&0&0&0\\
0&\e&0&0\\
0&0&u+2c_a+2s_a+1 +2{\bf S}^z&2{\bf S}^-\\
0&0&2{\bf S}^+&u+2c_a+2s_a+1-2{\bf S}^z\\
\end {array} \right ).\label{tl}
\eeq
which constitutes a realization of the Yang-Baxter algebra (\ref
{rtt-ttr}) when $\e$
tends to $0$.
The recent work of Frahm and Slavnov \cite{fs} confirms the existence of
such non-regular solutions by means of a projecting method. 

Finally, we would like to stress that here we have only considered the
case of Kondo impurities in these extended Hubbard models which are
based on the $sl(2)$ subalgebra of the bulk symmetry of the models. It
is of course possible to consider other boundary impurities
corresponding to different subalgebra embeddings such as $sl(1|1)$ for
the $gl(2|2),\,gl(3|1)$ cases or $sl(3)$ for the $gl(3|1),\,gl(4)$
models and even $gl(2|1)$ for $gl(3|1),\,gl(2|2)$. For the case of $t-J$
models such other types of integrable boundary impurities have been
studied in \cite{bf99}.

\vskip.3in
This work is supported by OPRS and UQPRS. 
Jon Links and Mark D.Gould are supported by an Australian Research Council.

\appendix

\section{Derivation of the non-c-number boundary K-matrices}

In this appendix, we sketch the procedure of solving the (${\bf Z}_2$-graded) 
RE for $K_-(u)$ . To describe integrable Kondo impurites coupled
with the one-dimensional supersymmetric extended Hubbard model open chain,it is reasonable to assume
that
\beq
K_-(u) =\left (
\begin {array} {cccc}
1&0&0&0\\
0&1&0&0\\
0&0&A(u)&B(u)\\
0&0&C(u)&D(u)\\
\end {array}  \right ).
\eeq
Throughout, we have omitted all the subscrips for brevity,  
reflecting  that the fermionic degrees of freedom do not occur, as it
should be for a magnetic impurity.
For the $R$-matrix (\ref{r1}), one may get from the RE (\ref {reflection1})
54 functional equations, of which 14 are identities. After some
algebraic analysis, together with the $su(2)$ symmetry,
we may assume that
\bea
A(u)&=&\a(u)+\b(u){\bf S}^z, ~~~B(u)=\b(u){\bf
S}^-,\no\\
C(u)&=&\b(u){\bf S}^+,~~~
D(u)=\a(u)-\b(u){\bf S}^z. \label {AK}
\eea
There are 10 equations automatically satisfied and 10 same equations ,
leaving only 20 equations
left to be solved
\bea
& &A(u_1)B(u_2)+B(u_1)D(u_2)=
A(u_2)B(u_1)+B(u_2)D(u_1),\no\\
& &C(u_1)A(u_2)+D(u_1)C(u_2)=
C(u_2)A(u_1)+D(u_2)C(u_1),\no\\
& &u_-(A(u_1)B(u_2)+B(u_1)D(u_2))=
u_+(B(u_1)-B(u_2)),\no\\
& &u_-(A(u_2)B(u_1)+B(u_2)D(u_1))=
u_+(B(u_1)-B(u_2)),\no\\
& &u_-(C(u_1)A(u_2)+D(u_1)C(u_2))=
u_+(C(u_1)-C(u_2)),\no\\
& &u_-(C(u_2)A(u_1)+D(u_2)C(u_1))=
u_+(C(u_1)-C(u_2)),\no\\
& &u_-(A(u_1)A(u_2)+B(u_1)C(u_2)-1)=
u_+(A(u_1)-A(u_2)),\no\\
& &u_-(A(u_2)A(u_1)+B(u_2)C(u_1)-1)=
u_+(A(u_1)-A(u_2)),\no\\
& &u_-(C(u_1)B(u_2)+D(u_1)D(u_2)-1)=
u_+(D(u_1)-D(u_2)),\no\\
& &u_-(C(u_2)B(u_1)+D(u_2)D(u_1)-1)=
u_+(D(u_1)-D(u_2)),\no\\
& &2u_-(A(u_1)B(u_2)+B(u_1)D(u_2))=
2u_+(D(u_2)B(u_1)-D(u_1)B(u_2))+u_+u_-(D(u_2)B(u_1)-B(u_1)D(u_2)),\no\\
& &2u_-(A(u_2)B(u_1)+B(u_2)D(u_1))=
2u_+(B(u_1)A(u_2)-B(u_2)A(u_1))+u_+u_-(B(u_1)A(u_2)-A(u_2)B(u_1)),\no\\
& &2u_-(C(u_1)A(u_2)+D(u_1)C(u_2))=
2u_+(A(u_2)C(u_1)-A(u_1)C(u_2))
+u_+u_-(A(u_2)C(u_1)-C(u_1)A(u_2)),\no\\
& &2u_-(C(u_2)A(u_1)+D(u_2)C(u_1))=
2u_+(C(u_1)D(u_2)-C(u_2)D(u_1))
+u_+u_-(C(u_1)D(u_2)-D(u_2)C(u_1)),\no\\
& &2u_-(A(u_2)A(u_1)+B(u_2)C(u_1)-C(u_1)B(u_2)-D(u_1)D(u_2))\no\\
& &~~~~~~~~~~~~~~~~~~~~=2u_+(A(u_1)D(u_2)-A(u_2)D(u_1))
-u_+u_-(B(u_2)C(u_1)-C(u_1)B(u_2)),\no\\
& &2u_-(A(u_1)A(u_2)+B(u_1)C(u_2)-C(u_2)B(u_1)-D(u_2)D(u_1))\no\\
& &~~~~~~~~~~~~~~~~~~~~=2u_+(D(u_2)A(u_1)-D(u_1)A(u_2))
-u_+u_-(B(u_1)C(u_2)-C(u_2)B(u_1)),\no\\
& &2u_-(A(u_1)B(u_2)+B(u_1)D(u_2))
+u_+u_-(A(u_1)B(u_2)-B(u_2)A(u_1))\no\\
& &~~~~~~~~~~~~~~=2u_+(A(u_2)B(u_1)-A(u_1)B(u_2))
+4(A(u_2)B(u_1)+B(u_2)D(u_1)-A(u_1)B(u_2)-B(u_1)D(u_2)),\no\\
& &2u_-(A(u_2)B(u_1)+B(u_2)D(u_1))
+u_+u_-(B(u_2)D(u_1)-D(u_1)B(u_2))\no\\
& &~~~~~~~~~~~~~~=2u_+(B(u_1)D(u_2)-B(u_2)D(u_1))
+4(A(u_1)B(u_2)+B(u_1)D(u_2)-A(u_2)B(u_1)-B(u_2)D(u_1)),\no\\
& &2u_-(C(u_1)A(u_2)+D(u_1)C(u_2))
+u_+u_-(D(u_1)C(u_2)-C(u_2)D(u_1))\no\\
& &~~~~~~~~~~~~~~=2u_+(D(u_2)C(u_1)-D(u_1)C(u_2))
+4(C(u_2)A(u_1)+D(u_2)C(u_1)-C(u_1)A(u_2)-D(u_1)C(u_2)),\no\\
& &2u_-(C(u_2)A(u_1)+D(u_2)C(u_1))
+u_+u_-(C(u_2)A(u_1)-A(u_1)C(u_2))\no\\
& &~~~~~~~~~~~~~~=2u_+(C(u_1)A(u_2)-C(u_2)A(u_1))
+4(C(u_1)A(u_2)+D(u_1)C(u_2)-C(u_2)A(u_1)-D(u_2)C(u_1)),
\eea
with $u_+=u_1+u_2,u_-=u_1-u_2$.
Substituting (\ref {AK}) into these equations,
we find that all these equations are
reduced to the following three equations
\bea
u_+(\a(u_1)-\a(u_2))&=&u_-(-1+\a(u_1)\a(u_2)+s(s+1)\b(u_1)\b(u_2)),\no\\
u_+(\b(u_1)-\b(u_2))&=&u_-(\a(u_1)\b(u_2)+\a(u_2)\b_(u_1)-\b(u_1)\b(u_2)),\no\\
2u_+(\a(u_2)\b(u_1)-\a(u_1)\b(u_2))&=&2 u_-(\a(u_1)\b(u_2)+\a(u_2)\b(u_1))-
u_-(u_++2)\b(u_1)\b(u_2)).
\eea
Taking the limit $u_1 \rightarrow u_2$,these equations become
\bea
\frac{d\a(u)}{du}&=&\frac{1}{2u}(-1+{\a(u)}^2+s(s+1){\b(u)}^2),\no\\
\frac{d\b(u)}{du}&=&\frac{1}{2u}(2\a(u)\b(u)-{\b(u)}^2),\no\\
\a(u)\frac{d\b(u)}{du}-\b(u)\frac{d\a(u)}{du}&=&
\frac{1}{2u}(2\a(u)\b(u)-(u+1){\b(u)}^2).
\label{A1}
\eea
Solving the first two equations, we have
\beq
\a(u)=\frac{(c_1c_2-u^2)(2s+1)+(c_2-c_1)u}
{(2s+1)(c_1-u)(c_2-u)},~~~~~~~~
\b(u)=\frac{2(c_2-c_1)u}
{(2s+1)(c_1-u)(c_2-u)},\label{A3}
\eeq
where $c_1$ and $c_2$ are integration constants. Substituting these 
Results into the third equation in (\ref{A1}),we may establish
a relation between $c_1 $ and $ c_2$:
$c_2=c_1-4s-2$.
This is nothing but the non-c-number
boundary $K$ matrix (\ref {k-1}) (after a redefinition of the 
constant:  
$c_1 \rightarrow 2c+4s+2$).

A similar construction also works for the quantum $R$ matrix (\ref
{r2}) and (\ref{r3}).

\end{document}